\documentclass[11pt,a4paper]{article}
\pdfoutput=1
\usepackage{jheppub}

\usepackage{color}
\usepackage{amsmath}
\usepackage{pifont}   
\usepackage{verbatim}       
\usepackage{acronym} 
    
\usepackage{amsfonts}
\usepackage{amssymb}
\usepackage{mathrsfs}
\usepackage{mathtools}
\usepackage{graphicx}
\usepackage{multirow} 
\usepackage{slashed} 
\usepackage{array}

\usepackage{graphicx}
\usepackage{epsfig}
\usepackage{lscape}
\usepackage{rotating}
\usepackage{epsf}
\usepackage{bm}
\usepackage{booktabs}
\usepackage{array}
\usepackage{caption}
\usepackage{subcaption}
\usepackage[utf8]{inputenc}
\usepackage{float}

\usepackage{multirow}
\usepackage{array,booktabs,colortbl,multirow}
\usepackage{colortbl,xcolor,colordvi,color}
\usepackage{rotating}
\allowdisplaybreaks

\usepackage{multicol}

\title{Dark matter semi-annihilation for inert scalar multiplets}

\author[a]{Hugues Beauchesne}
\author[b,a]{and Cheng-Wei Chiang}

\affiliation[a]{Physics Division, National Center for Theoretical Sciences,\\ Taipei 10617, Taiwan}
\affiliation[b]{Department of Physics and Center for Theoretical Physics, National Taiwan University, \\ Taipei 10617, Taiwan}

\emailAdd{beauchesneh@phys.ncts.ntu.edu.tw, chengwei@phys.ntu.edu.tw}

\abstract{Dark matter semi-annihilation is a process through which two dark matter candidates annihilate to a single dark matter particle and a non-dark matter particle. Such processes are common when the symmetry stabilizing the dark matter differs from $\mathbb{Z}_2$ and can lead to qualitatively different phenomenology. In this work, we study the viability of semi-annihilation models including one or two inert multiplets. For one multiplet, we show that there does not exist any viable model in which semi-annihilation is efficient. For two multiplets, semi-annihilation can be efficient, but the number of viable and technically natural models is limited. We then perform a detailed study of the most promising model, showing that the correct relic abundance can be obtained for a wide range of masses.}

\begin{document}

\maketitle

\section{Introduction}\label{Sec:Intro}
The preferred explanation for dark matter abundance has long been the standard freeze-out of Weakly Interacting Massive particles (WIMP). In this scenario, two dark matter particles of masses around the electroweak (EW) scale annihilate into Standard Model (SM) particles until the corresponding rate drops below the expansion rate of the Universe. Although this can naturally explain the observed dark matter abundance, standard WIMP thermal freeze-out is now under severe experimental constraints.

A potential explanation for dark matter that avoids many of the problems of standard thermal freeze-out is semi-annihilation~\cite{DEramo:2010keq}. In this mechanism, two dark matter candidates annihilate to a single dark matter particle and another particle that is not part of the dark matter content. These processes are common in models in which the dark matter stability is explained by a symmetry other than the common $\mathbb{Z}_2$ symmetry. Semi-annihilation processes have the advantage that they, by their very nature, do not directly lead to direct detection signals. This generally results in a qualitatively different allowed parameter space. Previous works on semi-annihilation include Refs.~\cite{Hambye:2008bq, Batell:2010bp, Belanger:2012vp, Ivanov:2012hc, Belanger:2014bga, Cai:2016hne, Arcadi:2017vis, Kamada:2017gfc, Queiroz:2019acr, Yaguna:2019cvp, Ghosh:2020lma, Belanger:2020hyh, Belanger:2021lwd, Yaguna:2021rds, DiazSaez:2022nhp, Belanger:2022qxt, Bandyopadhyay:2022tsf, Cheng:2022hcm, BasiBeneito:2022qxd, Guo:2023kqt, Benincasa:2023vyp}.

Some of the most studied dark matter candidates are scalar multiplets.\footnote{Throughout this article, by ``multiplet'' we refer to the $SU(2)_L$ multiplets that are color-neutral under $SU(3)_C$.} Multiplets considered in previous studies include doublets~\cite{Cirelli:2005uq, Gustafsson:2007pc, Hambye:2009pw, Garcia-Cely:2015khw, Chowdhury:2016mtl, Jangid:2020qgo, Jueid:2020rek, Bottaro:2022one}, triplets~\cite{Cirelli:2005uq, Hambye:2009pw, Araki:2011hm, Josse-Michaux:2012oqz, YaserAyazi:2014jby, Chowdhury:2016mtl, Khan:2016sxm, Lu:2016dbc, Kadota:2018lrt, Jangid:2020qgo, Bottaro:2021snn, Bottaro:2022one}, quadruplets~\cite{Cirelli:2005uq, Hambye:2009pw, AbdusSalam:2013eya, Cai:2017wdu, Zeng:2019tlw, Chowdhury:2016mtl, Bottaro:2022one}, quintuplets~\cite{Cirelli:2005uq, Cirelli:2007xd, Hambye:2009pw, Earl:2013jsa, Garcia-Cely:2015dda, Cai:2017fmr, Kadota:2018lrt, Chao:2018xwz, Bottaro:2021snn, Bottaro:2022one}, sextuplets~\cite{Cirelli:2005uq, Hambye:2009pw, Earl:2013jsa, Earl:2013fpa, Logan:2016ivc, Liu:2020dok, Bottaro:2022one}, septuplets~\cite{Cirelli:2005uq, Cirelli:2007xd, Hambye:2009pw, Cai:2012kt, Earl:2013jsa, Garcia-Cely:2015dda, Cai:2015kpa, Kadota:2018lrt, Chao:2018xwz, Bottaro:2021snn} and octuplets~\cite{Earl:2013jsa, Earl:2013fpa, Logan:2016ivc, Bottaro:2022one}. Combinations of different multiplets were also considered in Refs.~\cite{Fischer:2011zz, Fischer:2013hwa, Cheung:2013dua, AbdusSalam:2013eya, Kakizaki:2016dza, Cai:2017wdu, Liu:2017gfg, Cai:2018nob, Filimonova:2018qdc}. What almost all of these papers have in common is the assumption that dark matter is stabilized by a $\mathbb{Z}_2$ symmetry, which means that the dark matter abundance is set by the annihilation of dark matter particles to SM particles and not semi-annihilation.

In this paper, we study the possibility of semi-annihilation for inert multiplets. More precisely, we seek to determine which models of inert multiplets can lead to efficient semi-annihilation, while at the same time avoiding stable charged particles in a technically natural way and being compatible with known constraints. Models with one or two inert multiplets are considered. The most promising model is then examined in more detail by computing the dark matter density, studying various constraints, and mapping the region of parameter space compatible with the measured relic abundance.

We find the following results. Models with a single inert multiplet cannot lead to efficient semi-annihilation. For two multiplets, there exists a category of models that can lead to efficient semi-annihilation and no stable charged particles even at tree level. It also satisfies one of two possible conditions to suppress direct detection signal for inert scalar multiplets, i.e., the dark matter candidate being made up mostly of odd-dimensional multiplets of zero weak hypercharge or being almost pure CP-even or CP-odd scalars. Other models with two inert multiplets and efficient semi-annihilation exist, but cannot satisfy all of these properties simultaneously. For smaller multiplets, the existence of semi-annihilation processes allows multiplets to account for the correct relic abundance for masses varying by more than an order of magnitude. The allowed relative range shrinks as the size of the multiplets increases. Some combinations of gauge numbers can also lead to a strongly suppressed signal from indirect detection experiments.

The paper is organized as follows. Sec.~\ref{Sec:OneMultiplet} presents the case of a single multiplet and explains why it fails. Sec.~\ref{Sec:TwoMultiplets} considers the case of two multiplets. The model that we will focus on is presented in Sec.~\ref{Sec:Model}. The details of the dark matter abundance computation are given in Sec.~\ref{Sec:DarkMatterAbundanceComputation}. The procedure through which constraints are applied is explained in Sec.~\ref{Sec:Constraints}. Plots of the allowed parameter space are shown in Sec.~\ref{Sec:Results}. Concluding remarks are given in Sec.~\ref{Sec:Conclusion}, including a discussion of indirect detection signals. Some technical details about $SU(2)$ tensors are presented in Appendix~\ref{Sec:SU2}.

\section{One multiplet case}\label{Sec:OneMultiplet}
We begin by introducing the only unexcluded model that allows for semi-annihilation with only one multiplet. We then explain why it fails and what a viable model would require.

\subsection{Model}\label{sSec:OneMultipletModel}
Consider a complex scalar multiplet $\phi$ of dimension $n$ and weak hypercharge 0. The relevant renormalizable potential is
\begin{equation}\label{eq:Potential0A}
    V = V_0 + V_A + V_B + V_C,
\end{equation}
where
\begin{equation}\label{eq:Potential0B}
  \begin{aligned}
      V_0 &= -\mu^2 |H|^2 + \lambda_0 |H|^4 + m^2 |\phi|^2, \quad\quad
    & V_A &= \lambda_1 A_{abc} \phi^a \phi^b \phi^c + \text{h.c.},\\
      V_B &= \sum_{r=1}^\alpha \lambda_2^r B^r_{abcd} (H^a)^\dagger H^b (\phi^c)^\dagger \phi^d,
    & V_C &=  \sum_{r=1}^\beta \lambda_3^r C^r_{abcd} (\phi^a)^\dagger (\phi^b)^\dagger \phi^c \phi^d,
  \end{aligned}
\end{equation}
where the mass parameters $\mu^2$ and $m^2 > 0$, $\lambda_1$ has the dimension of mass, and $\lambda_{2,3}$ are dimensionless.
The indices represented by Latin letters besides $r$ correspond to $SU(2)$ indices and are summed over. The invariant tensors of the potential are presented in Appendix~\ref{sSec:SU2OneMultiplet}. The $A_{abc}$ term is non-zero for $n \in \{1, 5, 9, ...\}$. The upper limits on the sums are given by
\begin{equation}\label{eq:UpperLimits1}
  \begin{aligned}
    \alpha &= \text{min}\left(2, n\right), & \beta &= \left\lfloor\frac{n + 1}{2}\right\rfloor,
  \end{aligned}
\end{equation}
where $\lfloor x \rfloor$ means $x$ rounded down to the closest integer. The potential respects a $\mathbb{Z}_3$ symmetry under which only $\phi$ transforms in the following way:
\begin{equation}\label{eq:SymmetryV1}
  \phi \to e^{2\pi i/3}\phi.
\end{equation}
Unless $\mathbb{Z}_3$ is spontaneously broken, the lightest component of $\phi$ will be stable.

\subsection{Comments}\label{sSec:OneMultipletComments}
We now comment on the model and explain why it fails as a viable semi-annihilation model.

Besides the standard Higgs potential and multiplet masses, the potential consists of three types of terms. This separation will also apply to all other models defined in this work. Type $A$ terms contain exactly three multiplets and in this potential take the form $\lambda_1 \phi^3 + \text{h.c.}$ Without such terms, semi-annihilation would not be possible. Type $B$ terms are those that involve two $\phi$'s and Higgs doublets. They are responsible for splitting the masses of the components of the multiplet and also contribute to certain semi-annihilation processes. Type $C$ terms are dark matter self-interactions. Their exact details will not affect any computations in this paper.\footnote{The $C$ terms would however affect vacuum stability and global minima requirements. Since this is somewhat orthogonal to the rest of this work, we leave a detailed study of this for future work.}

This model has the nice property that it does not lead to any interactions between the neutral component of $\phi$ and the $Z$ boson at tree level. This makes it far more likely to be consistent with the very stringent bounds from direct detection experiments. However, the model suffers from two problems, which will prove challenging for inert multiplet semi-annihilation in general. 

First, the model suffers from a small naturalness problem. The potential includes a term that can be written as
\begin{equation}\label{eq:Splitting}
  \lambda_3^2 H^\dagger \tau^a H \phi^\dagger T^a \phi,
\end{equation}
where $\tau^a$ and $T^a$ are the appropriate $SU(2)$ generators. Besides small radiative corrections, this is the only source of mass splitting between the components of the multiplet. It leads to the lightest component of $\phi$ being either the one with the most positive or most negative value of $T^3$, depending on the sign of $\lambda_3^2$. Unless $\phi$ is a singlet, this results in the dark matter being electrically charged, which is not phenomenologically viable. Of course, the coefficient of this operator can be set to zero, but it is not technically natural as it does not correspond to a point of enhanced symmetry. In a more general case, this problem could be solved in three ways. First, this term is zero if the multiplet is real. This is often incompatible with the existence of symmetries allowing for semi-annihilation. Second, the multiplet can have a non-zero hypercharge. This can be done via $|Y| = (n - 1)/2$, resulting in the neutral state being the one with the largest or smallest $T^3$, or via $Y = -1/2$, in which case a term of the form $\lambda H^2 \phi^2 +\text{h.c.}$ can be included. Third, the multiplet could mix with other fields. This will require the presence of another multiplet. See Refs.~\cite{Earl:2013jsa, Earl:2013fpa} for enlightening discussions on the subject. In the end, this problem is more esthetical and is not sufficient to exclude the model. It is also not that difficult to have the lightest particle be neutral for scalars in the TeV or tens of TeV range when loop corrections are taken into account.

Second, the model suffers from a very serious practical problem that removes most of its appeal: It does not lead to efficient semi-annihilation. The reason is as follows. Two types of semi-annihilation processes are possible for this model:
\begin{enumerate}
  \item $\phi^a \phi^b \to (\phi^c)^\dagger\;\; h$,
  \item $\phi^a \phi^b \to (\phi^c)^\dagger\;\; B$,
\end{enumerate}
where $h$ is the Higgs boson and $B$ denotes a gauge boson $A$, $Z$, $W^+$ or $W^-$. The conjugate processes are of course also possible.

For the first process, the cross section is vanishingly small. The reason is that, if electroweak symmetries were unbroken, this process would be forbidden by conservation of both isospin and weak hypercharge. The cross section must then be suppressed by some power of $v/m$, where $v$ is the Higgs vacuum expectation value (VEV). Since electroweak multiplets must typically have a mass in the TeV range or higher to reproduce the correct dark matter abundance, this results in very inefficient processes of type 1 for this model.

For the second process, consider the limit in which $v$ is zero. The corresponding cross section is
\begin{equation}\label{eq:CS1}
  \sigma = \frac{1}{64\pi s p_A^2} \int_{t_1}^{t_0} |M|^2 dt,
\end{equation}
with $p_A$ being the center-of-mass three-momentum of one of the incoming particles and
\begin{equation}\label{eq:CS2}
  \int_{t_1}^{t_0} |M|^2 dt \propto \frac{-3\sqrt{s(s- 4m^2)} + 2(s + 2m^2) \text{tanh}^{-1}\left(\sqrt{\frac{s - 4m^2}{s}}\right)}{s - m^2} \approx \frac{\Delta s ^{5/2}}{90m^5},
\end{equation}
where the approximation is at leading order in $\Delta s = s - 4m^2$. A short computation shows that this leads to $\sigma \beta \propto \beta^4$, where $\beta$ is the M{\o}ller velocity, and therefore a thermally-averaged cross section $\langle \sigma \beta \rangle \propto T^2$, where $T$ is the temperature. When the VEV of the Higgs is taken into account, some of these processes are not quite $d$-wave, but the leading terms in the temperature expansion of the thermally averaged cross section are suppressed by powers of $v/m$, resulting in semi-annihilation processes of type 2 being mostly irrelevant. If one considers two multiplets $\phi_1$ and $\phi_2$, one could include in the potential a term of the form $\lambda \phi_1^2 \phi_2 + \text{h.c.}$ In the limit of zero Higgs VEV, the resulting semi-annihilation processes of type 2 would be $p$-wave if the incoming particles are distinct or $d$-wave otherwise, resulting in the same problem as for a single multiplet.

As a side note, this $p$-wave or $d$-wave suppression is unsurprising. Assume the collision $\phi_i \phi_j \to \phi_k^\dagger A$, with $A$ being a massless gauge boson and $i$, $j$ and $k$ labelling the species. The gauge boson $A$ has two possible helicities: $\pm 1$. To conserve angular momentum, the amplitude must be at least $p$-wave. For $i \neq j$, the amplitude can be $p$-wave and this is indeed the result we observe. For $i = j$, the amplitude must be invariant under the interchange of the two incoming scalars and the amplitude therefore cannot be $p$-wave. This is why the amplitude is $d$-wave for identical scalars. The case of $p$-wave suppression was also noted for gluons and photons in Ref.~\cite{Cai:2016hne}, though the argument was different.

The only way to obtain efficient semi-annihilation for a renormalizable Lagrangian with only one multiplet would then be to include a term of the form $\lambda \phi^3 H + \text{h.c.}$ The resulting semi-annihilation of type 1 would be $s$-wave and would not be suppressed by any power of $v/m$. Unfortunately, this would only lead to stable charged particles, which is experimentally problematic and does not provide any dark matter candidate. This rules out the possibility of efficient semi-annihilation for only one inert multiplet. 

From the discussion of this section, we are left with few options for efficient semi-annihilation of inert multiplets. First, one could add new fields beyond inert multiplets. This opens a few possibilities, but is beyond the scope of this paper. Second, the potential could include a term in $\lambda_1 \phi_1^3 + \text{h.c.}$ and a term in $\lambda_2 \phi_1 \phi_2^\dagger H^\dagger + \text{h.c.}$ This would lead to semi-annihilation processes involving the Higgs boson. However, $\lambda_2$ is constrained by direct detection experiments to be small, which would limit the usefulness of the semi-annihilation processes. Finally, the only option left is to include in the potential a term of the form $\lambda \phi_i \phi_j \phi_k H + \text{h.c.}$ with $i$, $j$ and $k$ not all equal.

\section{Two multiplets case}\label{Sec:TwoMultiplets}
Armed with our experience from the one multiplet case, we now consider models of two inert multiplets semi-annihilation. We will try to determine which models lead to efficient semi-annihilation, are more likely to avoid the bounds from direct detection experiments, and can naturally avoid stable charged particles.

First, as explained in the last section, a model of two inert multiplets with efficient semi-annihilation must include in its potential a term of the form
\begin{equation}\label{eq:phi1phi1phi2}
  V \supset \lambda \phi_1^2 \phi_2 H + \text{h.c.}
\end{equation}
The multiplet $\phi_2$ has to be an even representation of $SU(2)$ and therefore a complex multiplet. For both $\phi_1$ and $\phi_2$ to be non-trivially charged under a stabilizing symmetry and thus have semi-annihilation, $\phi_1$ must also be complex. We will refer to the weak hypercharge of $\phi_1$ and $\phi_2$ as $Y_1$ and $Y_2$ and their size as $n_1$ and $n_2$, respectively. By trying every combination of fields, one can verify that there is only a finite number of renormalizable type $B$ terms compatible with Eq.~\eqref{eq:phi1phi1phi2}. Of these, only some have both $\phi_1$ and $\phi_2$ non-trivially charged under a stabilizing symmetry and do not necessarily contain a stable charged particle. Besides $\lambda H^\dagger H \phi_i^\dagger \phi_i$ and $\lambda H^\dagger \tau^a H \phi_i^\dagger T^a \phi_i$ which are always allowed, the possibilities are:
\begin{itemize}
  \item $V_{B_1} = \lambda \phi_2^2 (H^\dagger)^2 + \text{h.c.}$, with $Y_1 = -1/2\;$, $Y_2 = 1/2\;$ and $\mathbb{Z}_4$.
  \item $V_{B_2} = \lambda \phi_1 \phi_2^\dagger H^2 + \text{h.c.}$, with $Y_1 = -1/2\;$, $Y_2 = 1/2\;$ and $\mathbb{Z}_3$.
  \item $V_{B_3} = \lambda \phi_2^2 H^2 + \text{h.c.}$, with $Y_1 = 0\;$, $Y_2 = -1/2\;$ and $\mathbb{Z}_4$.
  \item $V_{B_4} = \lambda \phi_1 \phi_2^\dagger H^\dagger + \text{h.c.}$, with $Y_1 = 0\;$, $Y_2 = -1/2\;$ and $\mathbb{Z}_3$.
  \item No $B$ term, with $U(1)$.
\end{itemize}
The symmetries above are respected by the $A$- and $B$-type terms and correspond to:
\begin{itemize}
  \item $\mathbb{Z}_3$: $\phi_1 \to e^{2\pi i/3} \phi_1$ and $\phi_2 \to e^{2\pi i/3} \phi_2$.
  \item $\mathbb{Z}_4$: $\phi_1 \to i\phi_1$ and $\phi_2 \to -\phi_2$.
  \item $U(1)$: $\phi_1 \to e^{\pi\alpha i} \phi_1$ and $\phi_2 \to e^{-2\pi\alpha i} \phi_2$, with $\alpha \in \mathbb{R}$.
\end{itemize}
The potential can satisfy one and only one of these scenarios. These symmetries could be broken by type $C$ terms, but doing so would not help satisfy any of our goals. A mass term of the form $m^2 \phi_1^2 + \text{h.c.}$ would be incompatible with the existence of semi-annihilation processes and a term of the form $m^2 \phi_2^2 + \text{h.c.}$ would actually be zero.

Second, without requiring precise accidental cancellations, the only two ways to suppress the direct detection signal of a stable scalar are:
\begin{itemize}
  \item Split the masses of the CP-even and CP-odd parts of the neutral component,
  \item Have dark matter be mostly a component from an odd-dimensional multiplet of zero weak hypercharge.
\end{itemize}
We will require one of these conditions to be satisfied for all dark matter components. For $V_{B_1}$, the lightest component of $\phi_1$ will be stable and cannot meet these requirements. For $V_{B_2}$, the lightest mass eigenstate will be stable and cannot meet these requirements.

Third, we would like the model to avoid the presence of stable charged particles in a technically natural way. For $V_{B_3}$, the lightest component of $\phi_1$ will be stable and there is no technically natural way for it to be the neutral component at tree level unless it is a singlet. For $U(1)$, the lightest component of $\phi_1$ will be stable and can only meet the direct detection requirements if $Y_1 = 0$. In this case, there is no natural way for the neutral component of $\phi_1$ to be the lightest at tree level unless it is a singlet.

Ultimately, the case of $V_{B_4}$ is the only scenario that meets all of the requirements without requiring $\phi_1$ to be a singlet. It will be the focus of the rest of this work and will be presented in more detail in the next section. The cases of $V_{B_1}$, $V_{B_3}$ and $U(1)$ are also not strictly excluded. They must, nevertheless, rely on loop corrections to avoid stable charged particles, satisfy special conditions to meet the direct detection constraints, or both. The scenario of $V_{B_2}$ would, however, be very difficult to salvage.

\section{Model definition}\label{Sec:Model}
In this section, we present the model which will be the focus of the rest of this work.

Consider two complex scalar multiplets $\phi_1$ and $\phi_2$ of dimensions $n_1$ and $n_2$ and with weak hypercharge $Y_1 = 0$ and $Y_2 = -1/2$, respectively.  We will assume throughout this work that $|n_1 - n_2| = 1$. The relevant potential is:
\begin{equation}\label{eq:Potential1A}
    V = V_0 + V_A + V_B + V_C,
\end{equation}
where
\begin{equation}\label{eq:Potential1B}
  \begin{aligned}
    V_0 =& -\mu^2 |H|^2 + \lambda_0 |H|^4 + m_1^2 |\phi_1|^2 + m_2^2 |\phi_2|^2,\\
    V_A =& \lambda_1 A_{abcd} \phi_1^a \phi_1^b \phi_2^c H^d + \lambda_2 B_{abc}\phi_1^a \phi_1^b \phi_1^c + \text{h.c.},\\
    V_B =& \sum_{r=1}^\alpha \lambda_3^r C^r_{abcd} (H^a)^\dagger H^b (\phi_1^c)^\dagger \phi_1^d + \sum_{r=1}^\beta \lambda_4^r D^r_{abcd} (H^a)^\dagger H^b (\phi_2^c)^\dagger \phi_2^d\\
         & + [\lambda_5 E_{abc} \phi_1^a (\phi_2^b)^\dagger (H^c)^\dagger + \text{h.c.}],\\
    V_C =& \sum_{r=1}^\gamma \lambda_6^r F^r_{abcd} (\phi_1^a)^\dagger (\phi_1^b)^\dagger \phi_1^c \phi_1^d + \sum_{r=1}^\delta \lambda_7^r G^r_{abcd} (\phi_2^a)^\dagger (\phi_2^b)^\dagger \phi_2^c \phi_2^d\\
         & + \sum_{r=1}^\epsilon \lambda_8^r H^r_{abcd} (\phi_1^a)^\dagger (\phi_2^b)^\dagger \phi_1^c \phi_2^d,
  \end{aligned}
\end{equation}
with $\lambda_{2,5}$ having the dimension of mass and all the other $\lambda$'s being dimensionless.
The $SU(2)$ tensors are given in Appendix~\ref{sSec:SU2TwoMultiplets}. The conditions for the $SU(2)$ contractions to be non-zero are
\begin{equation}\label{eq:TensorNonZero4}
  \begin{aligned}
    A: \quad & n_2 \in \{2n_1,\; 2n_1 - 2\;, 2n_1 - 4,\; ...\},\\
    B: \quad & n_1 \in \{1, 5, 9,\; ...\},\\
    E: \quad & n_1 - n_2 \in \{-1, 1\}.\\
  \end{aligned}
\end{equation}
All other contractions are always non-vanishing. The upper limits on the sums are given by
\begin{equation}\label{eq:UpperLimits4}
  \begin{aligned}
    \alpha &= \text{min}\left(2, n_1\right), & \beta &= \text{min}\left(2, n_2\right), & \gamma &= \left\lfloor\frac{n_1 + 1}{2}\right\rfloor, &
    \delta &= \left\lfloor\frac{n_2 + 1}{2}\right\rfloor, & \epsilon &= \text{min}(n_1, n_2).
  \end{aligned}
\end{equation}
The potential respects a $\mathbb{Z}_3$ symmetry under which
\begin{equation}\label{eq:SymmetryV4}
  \phi_1 \to e^{2\pi i/3}\phi_1, \quad \phi_2 \to e^{2\pi i/3}\phi_2.
\end{equation}
Unless $\mathbb{Z}_3$ is spontaneously broken, the lightest mass eigenstate will be stable.

\section{Dark matter abundance computation}\label{Sec:DarkMatterAbundanceComputation}
In this section, we present our computation of the dark matter abundance. The computation is performed first under the assumption of unbroken electroweak symmetries and then under the assumption of broken electroweak symmetries. We will show that the two methods are consistent within the expected precision. Our implementation of Sommerfeld enhancement (SE) is also presented.

\subsection{Computation for unbroken electroweak symmetries}\label{sSec:EWunbroken}
To compute the dark matter abundance under the assumption of unbroken electroweak symmetries, we will consider scattering from one weak isospin eigenstate to another. This will simplify much of the formalism, especially concerning the Sommerfeld enhancement. We take inspiration from and expand on the results of Refs.~\cite{Hally:2012pu, Earl:2013jsa, Earl:2013fpa, Logan:2016ivc}.

Consider a set of fields $\phi_1$, $\phi_2$ \ldots forming a state of isospin numbers $J$ and $M$. We will refer to this state as $[\phi_1 \phi_2 \text{\ldots}]^M_J$. Scattering will always be from a state $J$ and $M$ to a state of identical $J'$ and $M'$. The states can easily be constructed by a combination of Clebsch-Gordan coefficients and charge conjugation matrices.

\subsubsection*{Cross sections}
We begin by presenting the cross sections necessary for computing the dark matter abundance. All $2\to 2$ cross sections can be decomposed as
\begin{equation}\label{eq:CSdecomposition}
  \begin{aligned}
  \sigma([P_A P_B]^M_J \to [P_C P_D]^{M'}_J) &=  \sigma_0([P_A P_B]_J \to [P_C P_D]^{}_J) \delta^{MM'}\\
                                             &= \frac{S_{CD}}{64 \pi s p_A^2} \hat{M}([P_A P_B]_J \to [P_C P_D]_J) \delta^{MM'},
  \end{aligned}
\end{equation}
where $S_{CD}$ is $1/2$ if $P_C$ and $P_D$ are identical and $1$ otherwise and $p_A$ is the center-of-mass three-momentum of $P_A$ given by
\begin{equation}\label{eq:p1}
  p_A = \frac{\sqrt{(s - (m_A + m_B)^2)(s - (m_A - m_B)^2)}}{2\sqrt{s}},
\end{equation}
where $s$ is the usual Mandelstam variable. For annihilation to Higgs pairs, the $\hat{M}$ coefficients are given by
\begin{equation}\label{eq:phiphiHHCS}
  \begin{aligned}
    \hat{M}([\phi_1^\dagger \phi_1]_0 \to [H^\dagger H]_0) &= f_1\left(s, m_1, m_2, \sqrt{2n_1}\lambda_3^1, \sqrt{\frac{n_1}{2}}|\lambda_5|^2\right),\\
    \hat{M}([\phi_2^\dagger \phi_2]_0 \to [H^\dagger H]_0) &= f_1\left(s, m_2, m_1, \sqrt{2n_2}\lambda_4^1, \frac{n_1}{\sqrt{2n_2}}|\lambda_5|^2\right),\\
    \hat{M}([\phi_1^\dagger \phi_1]_1 \to [H^\dagger H]_1) &= f_1\left(s, m_1, m_2, \sqrt{\frac{(n_1^2 - 1)n_1}{24}}\lambda_3^2, \frac{(n_1 - n_2)}{\sqrt{6}}\frac{\sqrt{n_1(n_1^2 - 1)}}{n_2}|\lambda_5|^2\right),\\
    \hat{M}([\phi_2^\dagger \phi_2]_1 \to [H^\dagger H]_1) &= f_1\left(s, m_2, m_1, \sqrt{\frac{(n_2^2 - 1)n_2}{24}}\lambda_4^2, \frac{(n_1 - n_2)}{\sqrt{6}}\sqrt{\frac{n_2^2 - 1}{n_2}}|\lambda_5|^2\right),
  \end{aligned}
\end{equation}
where
\begin{equation}\label{eq:f1}
  \begin{aligned}
  &f_1(s, m_A, m_B, C_A, C_B) =\\ 
  &\hspace{1.2cm}\sqrt{s(s - 4m_A^2)}\left(C_A^2 + \frac{C_B^2}{(m_A - m_B)^2 + m_B^2 s} \right) - 4 C_A C_B \text{tanh}^{-1}\left(\frac{\sqrt{s(s - 4 m_A^2)}}{s - 2m_A^2 + 2m_B^2}\right).
  \end{aligned}
\end{equation}
The mass of the Higgs boson has been neglected, as will be the case for all computations in this subsection. For annihilation to pairs of gauge bosons, the $\hat{M}$ coefficients are given by
\begin{equation}\label{eq:phiphigaugegauge}
  \begin{aligned}
    \hat{M}([\phi_1^\dagger \phi_1]_0 \to [W W]_0) &= f_2\left(s, m_1, \frac{(n_1^2 - 1)\sqrt{n_1}}{4\sqrt{3}}g^2 \right),\\
    \hat{M}([\phi_2^\dagger \phi_2]_0 \to [W W]_0) &= f_2\left(s, m_2, \frac{(n_2^2 - 1)\sqrt{n_2}}{4\sqrt{3}}g^2 \right),\\
    \hat{M}([\phi_1^\dagger \phi_1]_2 \to [W W]_2) &= f_2\left(s, m_1, \frac{1}{2}\sqrt{\frac{n_1(n_1^2 - 1)(n_1^2 - 4)}{30}}g^2 \right),\\
    \hat{M}([\phi_2^\dagger \phi_2]_2 \to [W W]_2) &= f_2\left(s, m_2, \frac{1}{2}\sqrt{\frac{n_2(n_2^2 - 1)(n_2^2 - 4)}{30}}g^2 \right),\\
    \hat{M}([\phi_2^\dagger \phi_2]_1 \to [W B]_1) &= f_2\left(s, m_2, \sqrt{\frac{n_2(n_2^2 - 1)}{12}}Y_2 g g' \right),\\
    \hat{M}([\phi_2^\dagger \phi_2]_0 \to [B B]_0) &= f_2\left(s, m_2, (Y_2 g')^2 \right),\\
  \end{aligned}
\end{equation}
where $g$ and $g'$ are respectively the $SU(2)_L$ and $U(1)_Y$ gauge couplings and
\begin{equation}\label{eq:f2}
  f_2(s, m, C) = \frac{8 C^2}{s}\left[(s + 4m^2)\sqrt{s(s - 4 m^2)} - 8m^2(s - 2m^2)\text{tanh}^{-1}\left(\sqrt{\frac{s - 4m^2}{s}} \right)\right].
\end{equation}
For semi-annihilation involving a Higgs boson, we will assume that the $B$ tensor is zero, $\lambda_2$ is small or $\lambda_5$ is small. The latter condition must realistically be respected to meet direct detection constraints. If at least one of these conditions is satisfied, then the $\hat{M}$ coefficients are given by
\begin{equation}\label{eq:phiphiphiH}
  \begin{aligned}
    \hat{M}\left([\phi_1 \phi_1]_{J_0} \to [\phi_2^\dagger H^\dagger]_{J_0}\right) &= \frac{4|\lambda_1|^2}{R}\frac{\sqrt{s - 4m_1^2}(s - m_2^2)}{\sqrt{s}},\\
    \hat{M}\left([\phi_1 \phi_2]_{J_1} \to [\phi_1^\dagger H^\dagger]_{J_1}\right) &= 4|\lambda_1|^2\left|C^{R'_1}_{n_1 n_2}\right|^2\frac{\sqrt{(s - (m_1 + m_2)^2)(s - (m_1 - m_2)^2)}(s - m_1^2)}{s},\\
    \hat{M}\left([\phi_1 \phi_2]_{J_2} \to [\phi_1^\dagger H^\dagger]_{J_2}\right) &= 4|\lambda_1|^2\left|C^{R'_2}_{n_1 n_2}\right|^2\frac{\sqrt{(s - (m_1 + m_2)^2)(s - (m_1 - m_2)^2)}(s - m_1^2)}{s},\\
  \end{aligned}
\end{equation}
where
\begin{equation}\label{eq:phiphiphiH2}
  \begin{aligned}
    R &= 4\left\lfloor\frac{n_2}{4}\right\rfloor + 1, \quad & R'_1 &= n_1 + 1,           \quad & R'_2 &= n_1 - 1,\\
    J_0 &= \frac{R - 1}{2},                           \quad & J_1 &= \frac{R'_1 - 1}{2}, \quad & J_2  &= \frac{R'_2 - 1}{2},
  \end{aligned}
\end{equation}
and 
\begin{equation}\label{eq:CRpn1n2}
  \left|C^{R'}_{n_1 n_2}\right|^2 = \frac{1}{R'}\left(\frac{1}{2} - \frac{n_2}{4n_1}(R' - n_1)(n_1 - n_2)(-1)^{\frac{n_1 - 1}{2}}\right).
\end{equation}
For this model, semi-annihilation processes involving gauge bosons are either $d$-wave or forbidden by conservation of isospin or weak hypercharge. We therefore neglect them. The processes $\phi_1 \phi_2^\dagger \to H W$ and $\phi_1 \phi_2^\dagger \to H B$ are also possible, but they are $p$-wave and their cross sections are proportional to $|\lambda_5|^2$, which is forced to be small. We also neglect them.

\subsubsection*{Boltzmann equation}
We now proceed to derive the Boltzmann equation for the combined abundance of $\phi_1$ and $\phi_2$. We first define the effective cross sections
\begin{equation}\label{eq:effCS1}
  \begin{aligned}
    \sigma(\phi_1 \phi_1 \to \text{SM SM})     =\quad & \frac{1}{2n_1^2}\sigma_0\left([\phi_1^\dagger \phi_1]_0 \to [H^\dagger H]_0\right) + \frac{3}{2n_1^2}\sigma_0\left([\phi_1^\dagger \phi_1]_1 \to [H^\dagger H]_1\right)\\
                                                    + & \frac{1}{2n_1^2}\sigma_0\left([\phi_1^\dagger \phi_1]_0 \to [W W]_0\right)         + \frac{5}{2n_1^2}\sigma_0\left([\phi_1^\dagger \phi_1]_2 \to [W  W]_2\right),\\
    \sigma(\phi_2 \phi_2 \to \text{SM SM})     =\quad & \frac{1}{2n_2^2}\sigma_0\left([\phi_2^\dagger \phi_2]_0 \to [H^\dagger H]_0\right) + \frac{3}{2n_2^2}\sigma_0\left([\phi_2^\dagger \phi_2]_1 \to [H^\dagger H]_1\right)\\
                                                    + & \frac{1}{2n_2^2}\sigma_0\left([\phi_2^\dagger \phi_2]_0 \to [W W]_0\right)         + \frac{5}{2n_2^2}\sigma_0\left([\phi_2^\dagger \phi_2]_2 \to [W W]_2\right)\\
                                                    + & \frac{1}{2n_2^2}\sigma_0\left([\phi_2^\dagger \phi_2]_0 \to [B B]_0\right)         + \frac{3}{2n_2^2}\sigma_0\left([\phi_2^\dagger \phi_2]_1 \to [W B]_1\right),\\
    \sigma(\phi_1 \phi_1 \to \phi_2\; \text{SM}) =\quad & \frac{R}{4n_1^2}\sigma_0\left([\phi_1 \phi_1]_{J_0} \to [\phi_2^\dagger H^\dagger]_{J_0}\right),\\
    \sigma(\phi_1 \phi_2 \to \phi_1\; \text{SM}) =\quad & \hspace{-0.3cm} \frac{R'_1}{2 n_1 n_2}\sigma_0\left([\phi_1 \phi_2]_{J_1} \to [\phi_1^\dagger H^\dagger]_{J_1}\right)
                                                      + \frac{R'_2}{2 n_1 n_2}\sigma_0\left([\phi_1 \phi_2]_{J_2} \to [\phi_1^\dagger H^\dagger]_{J_2}\right).\\
  \end{aligned}
\end{equation}

We define the number density per entropy density $s_E$ of $\phi_i$ as $Y_i$ and $Y = Y_1 + Y_2$. Their equilibrium distributions at temperature $T$ are given by 
\begin{equation}\label{eq:effCS2def}
 Y_i^\text{eq}(T) = \frac{g_im_i^2T}{2\pi^2 s_E}K_2(m_i/T),
\end{equation}
where $g_i = 2n_i$ is the number of internal degrees of freedom of $\phi_i$ and $K_n(x)$ are the modified Bessel functions of the second kind.

As long as $\lambda_5$ is not vanishingly small, $\phi_1$ and $\phi_2$ will remain in relative chemical equilibrium during freeze-out via interactions with gauge or Higgs bosons. We can then further define effective thermally averaged cross sections as
\begin{equation}\label{eq:effCS2}
  \begin{aligned}
    \langle\sigma \beta (\phi \phi \to \text{SM SM})\rangle     &= \left(\frac{Y_1^{\text{eq}}}{Y^{\text{eq}}}\right)^2\langle \sigma\beta(\phi_1 \phi_1 \to \text{SM SM}) \rangle
                                                                 + \left(\frac{Y_2^{\text{eq}}}{Y^{\text{eq}}}\right)^2\langle \sigma\beta(\phi_2 \phi_2 \to \text{SM SM}) \rangle,\\
    \langle\sigma \beta (\phi \phi \to \phi\; \text{SM})\rangle &= \left(\frac{Y_1^{\text{eq}}}{Y^{\text{eq}}}\right)^2\langle \sigma\beta(\phi_1 \phi_1 \to \phi_2 \;\text{SM}) \rangle
                                                                 + \frac{Y_1^{\text{eq}} Y_2^{\text{eq}}}{(Y^{\text{eq}})^2}\langle \sigma\beta(\phi_1 \phi_2 \to \phi_1 \;\text{SM}) \rangle.
  \end{aligned}
\end{equation}
The current dark matter number density per entropy density can then be computed by solving a single Boltzmann equation given by
\begin{equation}\label{eq:Boltzmann}
  \frac{dY}{dx} = -s_E\frac{dt}{dx}\left[\langle\sigma \beta (\phi \phi \to \text{SM SM})\rangle (Y^2 - (Y^{\text{eq}})^2) + \langle\sigma \beta (\phi \phi \to \phi\; \text{SM})\rangle Y(Y - Y^{\text{eq}})\right],
\end{equation}
where we define $x = m_1/T$ and
\begin{equation}\label{eq:dtdx}
  \frac{dt}{dx} = \sqrt{\frac{45}{4\pi^3}}\frac{g_\ast^{1/2}}{h_{\text{eff}}}\frac{M_{\text{Pl}}}{m_1T},\quad 
  g_\ast^{1/2} = \frac{h_{\text{eff}}}{g^{1/2}_{\text{eff}}}\left(1 + \frac{T}{3 h_{\text{eff}}}\frac{d h_{\text{eff}}}{dT}\right),
\end{equation}
with $g_{\text{eff}}$ and $h_{\text{eff}}$ respectively denoting the effective number of energy and entropy degrees of freedom and $M_{\text{Pl}}$ being the Planck mass. Once the present-day value of $Y$ has been obtained, the mass of the lightest eigenstate is computed, including one loop radiative corrections~\cite{Cirelli:2005uq}. These two quantities are then easily combined to obtain the dark matter relic abundance.

\subsubsection*{Sommerfeld enhancement}
Thermally averaged cross sections can be modified by the Sommerfeld enhancement because of the exchange of electroweak gauge bosons~\cite{Sommerfeld:1931qaf}. Being massless in the limit of unbroken electroweak symmetries, they mediate a Coulomb potential.

Consider two colliding particles $P_A$ and $P_B$ of weak isospins $J_A$ and $J_B$ and weak hypercharge $Y_A$ and $Y_B$, respectively. Assume they form a state of weak isospin $J$. Define~\cite{Strumia:2008cf}
\begin{equation}\label{eq:alpha}
  \alpha = \frac{g^2}{8\pi}\left[J(J + 1) - J_A(J_A + 1) - J_B(J_B + 1) \right] + \frac{(g')^2}{4\pi}Y_A Y_B \eta,
\end{equation}
where $\eta = -1$ for the collision of a particle and an antiparticle and $\eta = 1$ otherwise. The cross section is then enhanced by a multiplicative factor of~\cite{Harris:1957zza}
\begin{equation}\label{eq:SE}
  S_\text{SE} = -\frac{2\pi \alpha}{\beta}\frac{1}{1 - e^{\frac{2\pi \alpha}{\beta}}}.
\end{equation}
This enhancement is taken into account in the computation of the thermally averaged cross sections. If $\lambda_5$ were to be sufficiently large, a significant Sommerfeld enhancement via Higgs exchange could be possible. However, direct detection constraints on $\lambda_5$ limit this effect to be small and we neglect it.

\subsection{Computation for broken electroweak symmetries}\label{sSec:EWbroken}
We also perform the computation of the dark matter abundance when the VEV of the Higgs is set to its present-day value. Due to the presence of so many different masses, the amplitudes are cumbersome and computations are handled numerically.

Our procedure is as follows. Mass matrices are extracted from the Lagrangian and diagonalized. Interaction coefficients for mass eigenstates are then computed. The amplitudes for all annihilation and semi-annihilation processes involving gauge bosons or the Higgs boson are evaluated. The amplitudes are then used to compute the thermally averaged cross sections. The Boltzmann equation is solved numerically assuming all scalars maintain relative chemical equilibrium.

\subsection{Comparison of the methods}\label{sSec:Comparison}
\begin{figure}[t!]
    \centering
    \begin{subfigure}{0.49\textwidth}
        \centering
        \caption{}
        \includegraphics[width=\textwidth]{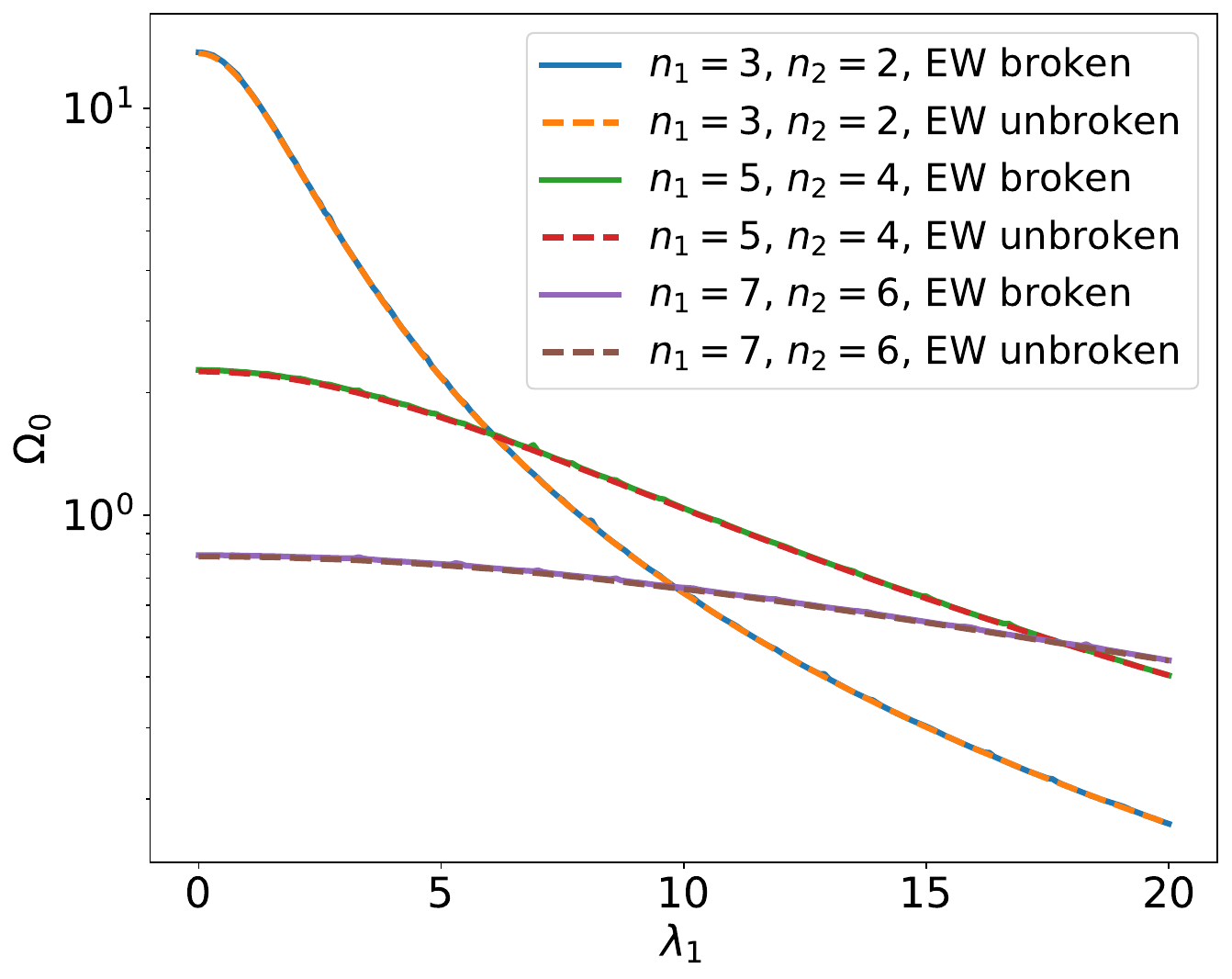}
        \label{fig:Validation}
    \end{subfigure}    
    \begin{subfigure}{0.49\textwidth}
        \centering
        \caption{}
        \includegraphics[width=\textwidth]{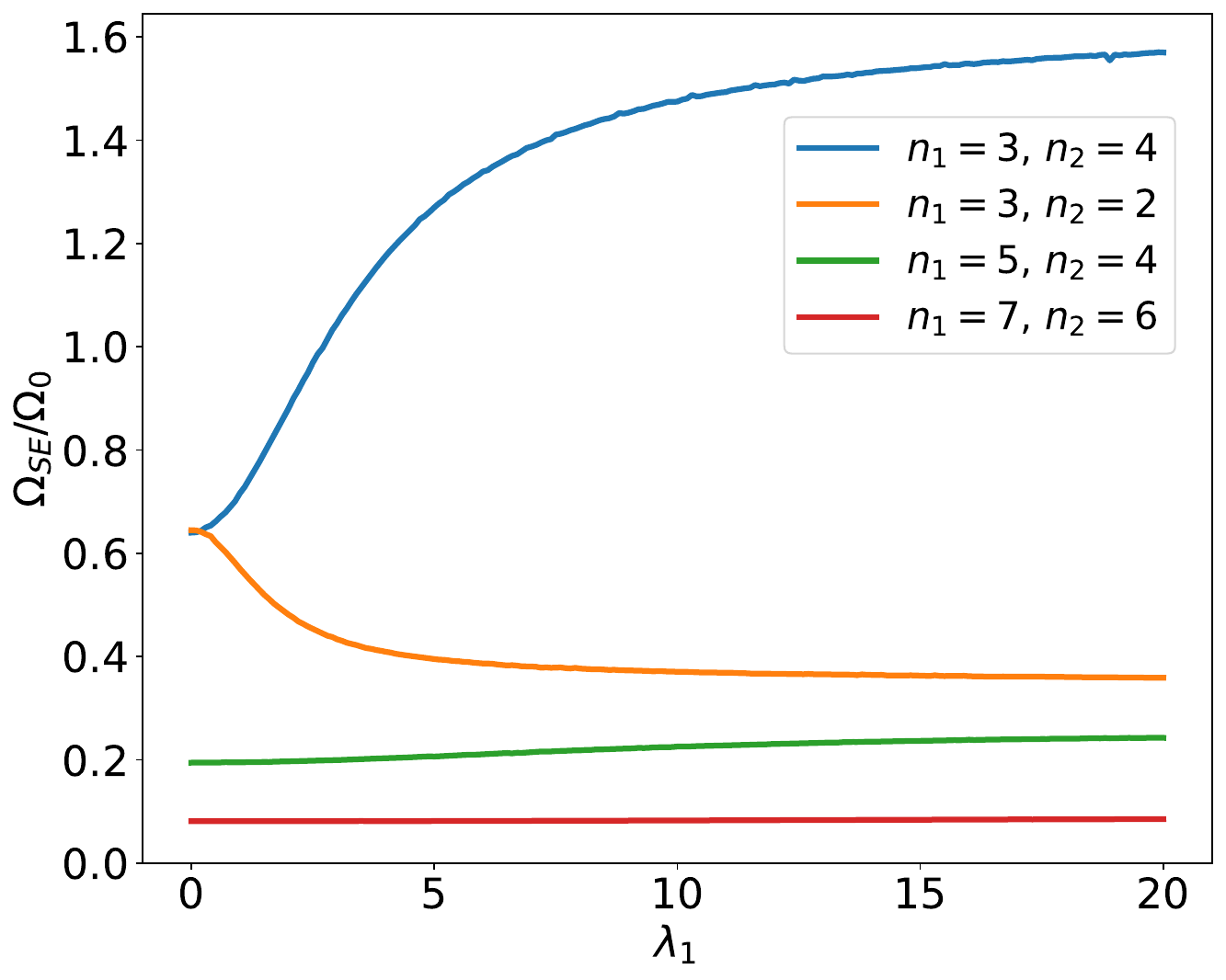}
        \label{fig:SE}
    \end{subfigure}    
    \caption{(a) Relic abundance $\Omega_0$ for different choices of $n_1$ and $n_2$ and for the methods of broken and unbroken electroweak symmetries. The Sommerfeld enhancement is not included in this case. The cases of $n_1$ and $n_2 = n_1 \pm 1$ are nearly identical and only $n_2 = n_1 - 1$ is shown. (b) Ratio of the relic abundance with the Sommerfeld enhancement $\Omega_{SE}$ and without $\Omega_0$. The case of $n_1 = 5$ and $n_2 = 6$ is nearly identical to $n_1 = 5$ and $n_2 = 4$ and is not included. For both plots, $m_1 = 10$~TeV, $m_2 = 11$~TeV and all other parameters are set to 0.}
    \label{fig:ValidationSE}
\end{figure}

We show in Fig.~\ref{fig:Validation} examples of the dark matter abundance computed using both methods. For these examples, the two methods give results consistent within about $1\%$. Such deviations are of the expected size of $\mathcal{O}(v^2/m_i^2)$, though part of these deviations are unavoidably due to numerical errors. Obtaining higher precision would require tracking the VEV of the Higgs during the freeze-out process, which is beyond the scope of this paper, and higher numerical precision. As can be seen in Fig.~\ref{fig:SE}, the Sommerfeld enhancement can substantially modify the dark matter abundance, especially for larger multiplets. Note also that semi-annihilation can in fact be made less efficient by the Sommerfeld enhancement, as is noticeably the case for $n_1 = 3$ and $n_2 = 4$. Since both methods give identical results to good precision, we will perform all computations in the limit of unbroken electroweak symmetries for the rest of this paper, including the Sommerfeld enhancement.

\section{Constraints}\label{Sec:Constraints}
In this section, we discuss our implementation of direct detection and unitarity constraints. Other constraints that were considered but ultimately not implemented are also discussed. Though a full study is beyond the scope of this work, we will discuss constraints from indirect detection in the conclusion.

\subsection{Direct detection experiments}\label{sSec:DirectDetection}
The lightest mass eigenstate $\hat{\phi}_1$ can in principle interact at tree level with both the $Z$ and Higgs bosons. We parameterize these interactions as
\begin{equation}\label{eq:LagrangianDD}
  \mathcal{L} \supset \left[-i \hat{C}_{11} Z_\mu \hat{\phi}_1^\dagger\partial^\mu \hat{\phi}_1 + \text{h.c.}\right] - h\hat{\Omega}_{11} \hat{\phi}_1^\dagger \hat{\phi}_1.
\end{equation}
The $\hat{C}_{11}$ and $\hat{\Omega}_{11}$ coefficients are obtained numerically and, when $\hat{\phi}_1$ is mostly the neutral component of $\phi_1$, respect to good approximation
\begin{equation}\label{eq:C11Omega11}
  \hat{C}_{11} \approx \frac{\sqrt{g^2 + g'{}^2}}{8} \frac{|\lambda_5 v|^2}{(m_2^2 - m_1^2)^2}, \quad
  \hat{\Omega}_{11} \approx -\frac{|\lambda_5|^2v}{2(m_2^2 - m_1^2)} + \lambda_3^1 v.
\end{equation}
Using the notation of Ref.~\cite{Hill:2014yxa}, the spin-independent per-nucleon cross section is
\begin{equation}\label{eq:DDCS}
  \sigma_{\text{SI}, N} = \frac{m_N^2 m_{\hat{\phi}_1}^2}{4\pi(m_N +  m_{\hat{\phi}_1})^2}\left[\sum_q \frac{2 g g_V^q}{c_W m_Z^2} \hat{C}_{11} F_1^{(N, q)}(0) - \frac{m_N}{m_h^2 m_{\hat{\phi}_1}}\frac{\hat{\Omega}_{11}}{v} f^{(0)}_{q, N} \right]^2,
\end{equation}
where $N$ is either a proton or a neutron, $c_W$ ($s_W$) is the cosine (sine) of the Weinberg angle, $v \approx 246$~GeV is the Higgs VEV, and $g_V^q = T^3_q/2 - Q_q s_W^2$. For the $Z$ exchange, we use the values $F_1^{(p, u)}(0) = F_1^{(n, d)}(0) = 2$, $F_1^{(p, d)}(0) = F_1^{(n, u)}(0) = 1$ and set all other $F_1^{(N, q)}(0)$ to 0. For Higgs exchange, we use $\sum_q f^{(0)}_{q, p} = 0.301$ and $\sum_q f^{(0)}_{q, n} = 0.307$~\cite{Ellis:2018dmb}. The values of $\sigma_{\text{SI}, N}$ are very close for the proton and neutron and we take their average. Loop corrections to $\sigma_{\text{SI}, N}$ were considered. However, a careful treatment reveals partial cancellations of the amplitudes such that they are safely below current detection limits for all multiplet sizes relevant to this work~\cite{Chen:2023bwg}. Constraints on $\sigma_{\text{SI}, N}$ are applied using the results of the LZ experiment from Ref.~\cite{LZ:2022lsv}.

\subsection{Unitarity}\label{sSec:Unitarity}
The parameter $\lambda_1$ is constrained by unitarity. Consider the process $\phi_1 \phi_1 \to \phi_2^\dagger H^\dagger$. In the limit of high energies, the amplitude is given in the basis of weak isospin eigenstates as
\begin{equation}\label{eq:AmplitudeUnitariy}
   M\left([\phi_1 \phi_1]_J^M \to [\phi_2^\dagger H^\dagger]_{J'}^{M'}\right) = \frac{2\lambda_1}{\sqrt{R}}\delta^{MM'} \delta_{JR}\delta_{J'R},
\end{equation}
where $R$ was defined in Eq.~\eqref{eq:phiphiphiH2}. It can be decomposed in terms of the Legendre polynomials $P_\ell(\cos\theta)$ as
\begin{equation}\label{eq:PartialWaves}
   M\left([\phi_1 \phi_1]_J^M \to [\phi_2^\dagger H^\dagger]_{J'}^{M'}\right) = 16\pi\sum_\ell (2\ell + 1) (a_\ell)^{MM'}_{JJ'} P_\ell(\cos\theta).
\end{equation}
This gives
\begin{equation}\label{eq:al}
  (a_0)^{MM'}_{JJ'} = \frac{\lambda_1}{8\pi\sqrt{R}}\delta^{MM'} \delta_{JR}\delta_{J'R}.
\end{equation}
In the basis of $[\phi_1 \phi_1]_J^M \to [\phi_2^\dagger H^\dagger]_{J'}^{M'}$, $(a_0)^{MM'}_{JJ'}$ can be written as a diagonal matrix. Taking into account the presence of two identical particles in the incoming state, unitarity requires the largest eigenvalue of this matrix in absolute value to be smaller than $1/\sqrt{2}$. This imposes the simple condition
\begin{equation}\label{eq:UnitarityFinal}
  |\lambda_1| \leq 4 \pi \sqrt{2 R}.
\end{equation}
As a side note, the fraction of $[\phi_1 \phi_1]_J^M$ states that contribute to this scattering scales roughly as $1/R$. A higher upper limit on $|\lambda_1|^2$ proportional to $R$ will partially compensate for this effect.

\subsection{Other constraints}\label{sSec:OtherConstraints}
Other constraints were considered but were found to be irrelevant and as such were not used in the final results. We computed the oblique parameters~\cite{Peskin:1990zt} using the results of Refs.~\cite{Beauchesne:2022fet, Beauchesne:2023bcy, Beauchesne:2023iyn}. All relevant constraints were found to be subdominant to either direct detection constraints or the requirement of having the stable component be neutral. Unitarity constraints on parameters besides $\lambda_1$ were studied using the results of Ref.~\cite{Beauchesne:2023iyn}, but were found to be satisfied in all regions of parameter space considered in this work. The Higgs signal strengths were also studied using the method of Ref.~\cite{Beauchesne:2023iyn}, but the scalars proved to be simply too heavy to have any significant impact on the Higgs properties.

\section{Results}\label{Sec:Results}
We present in this section some scans of the parameter space.

\begin{figure}[t!]
    \centering
    \begin{subfigure}{0.49\textwidth}
        \centering
        \caption{}
        \includegraphics[width=\textwidth]{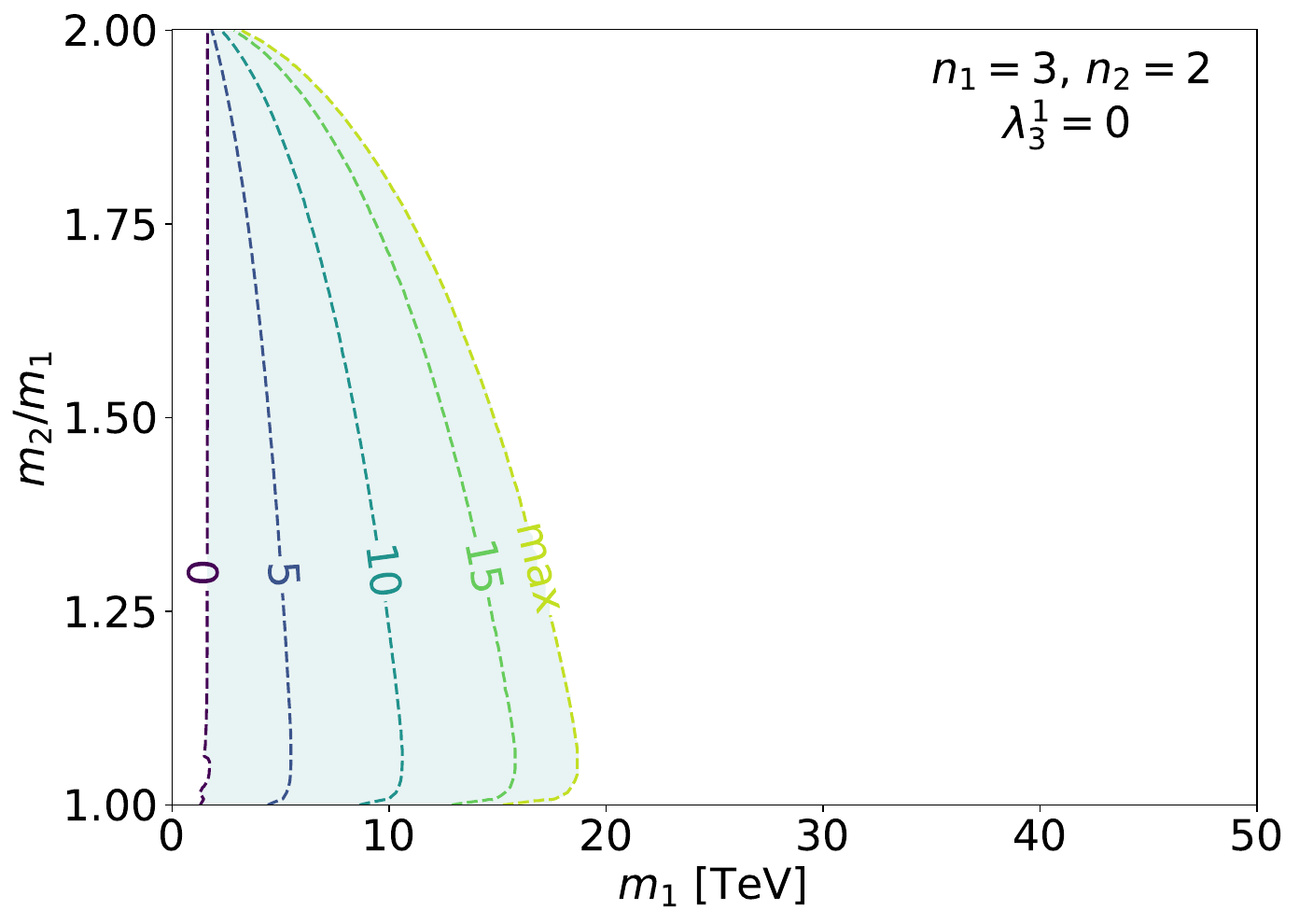}
        \label{fig:C32a}
    \end{subfigure}    
    \begin{subfigure}{0.49\textwidth}
        \centering
        \caption{}
        \includegraphics[width=\textwidth]{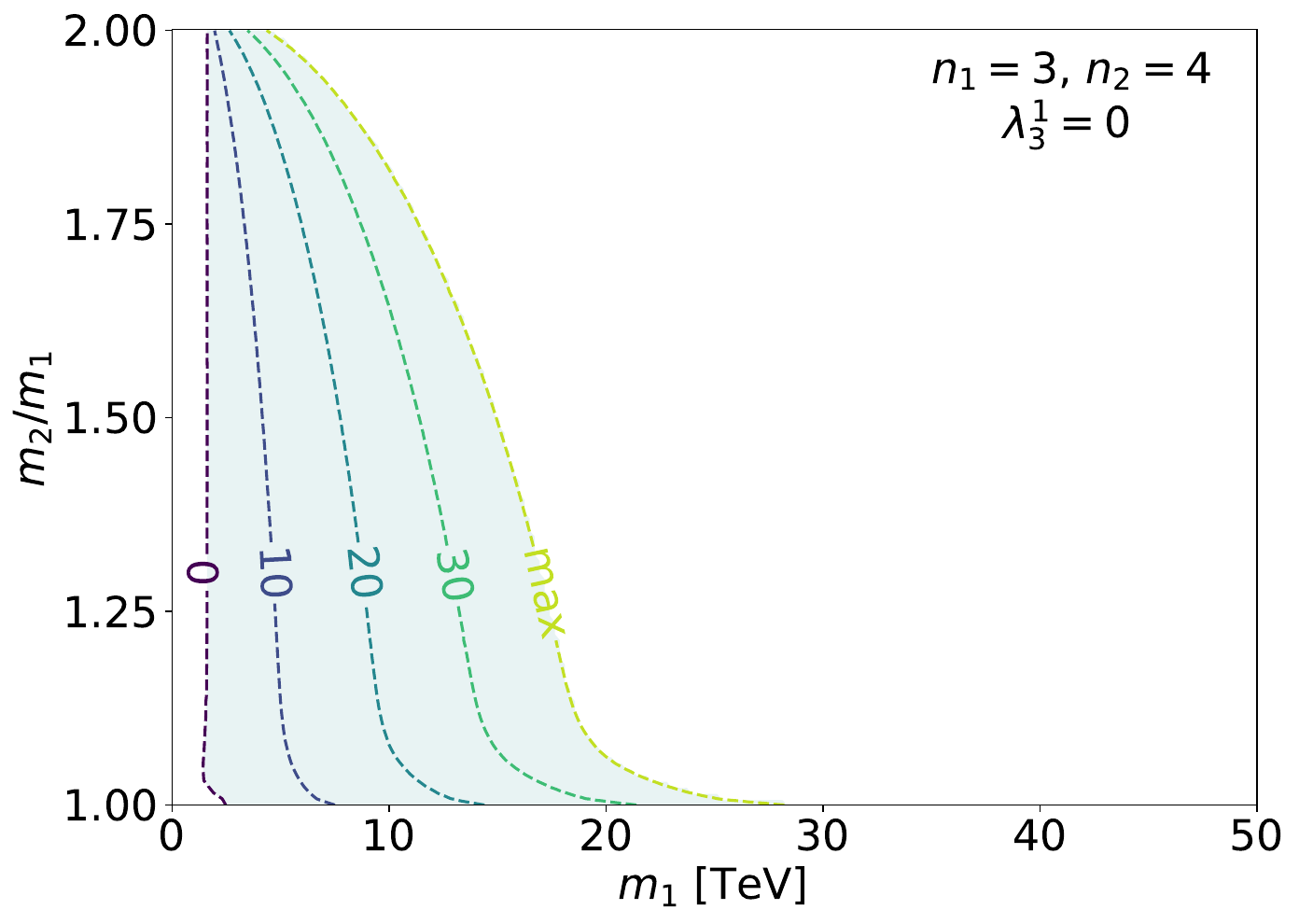}
        \label{fig:C34a}
    \end{subfigure}
    \begin{subfigure}{0.49\textwidth}
        \centering
        \caption{}
        \includegraphics[width=\textwidth]{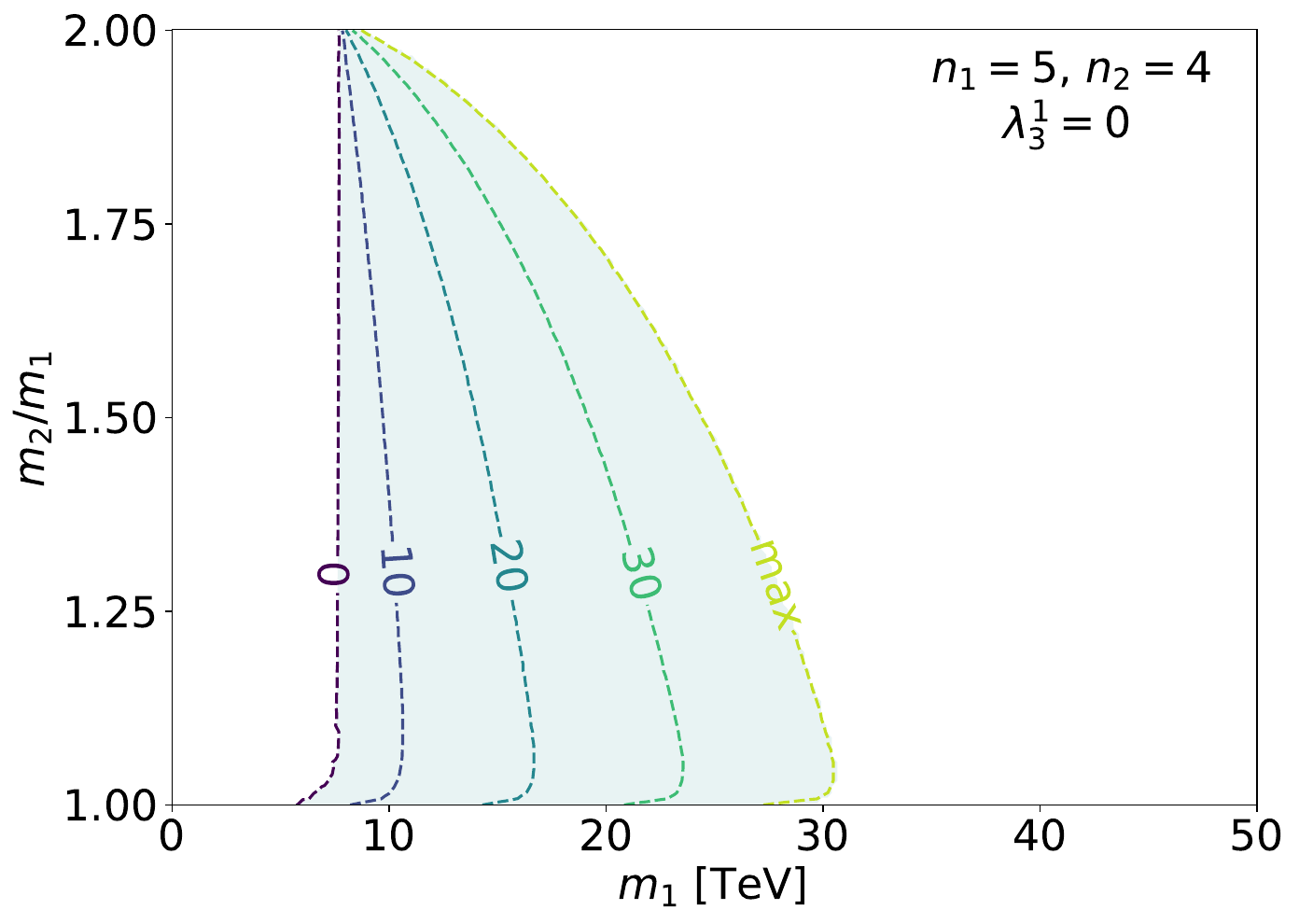}
        \label{fig:C54a}
    \end{subfigure}    
    \begin{subfigure}{0.49\textwidth}
        \centering
        \caption{}
        \includegraphics[width=\textwidth]{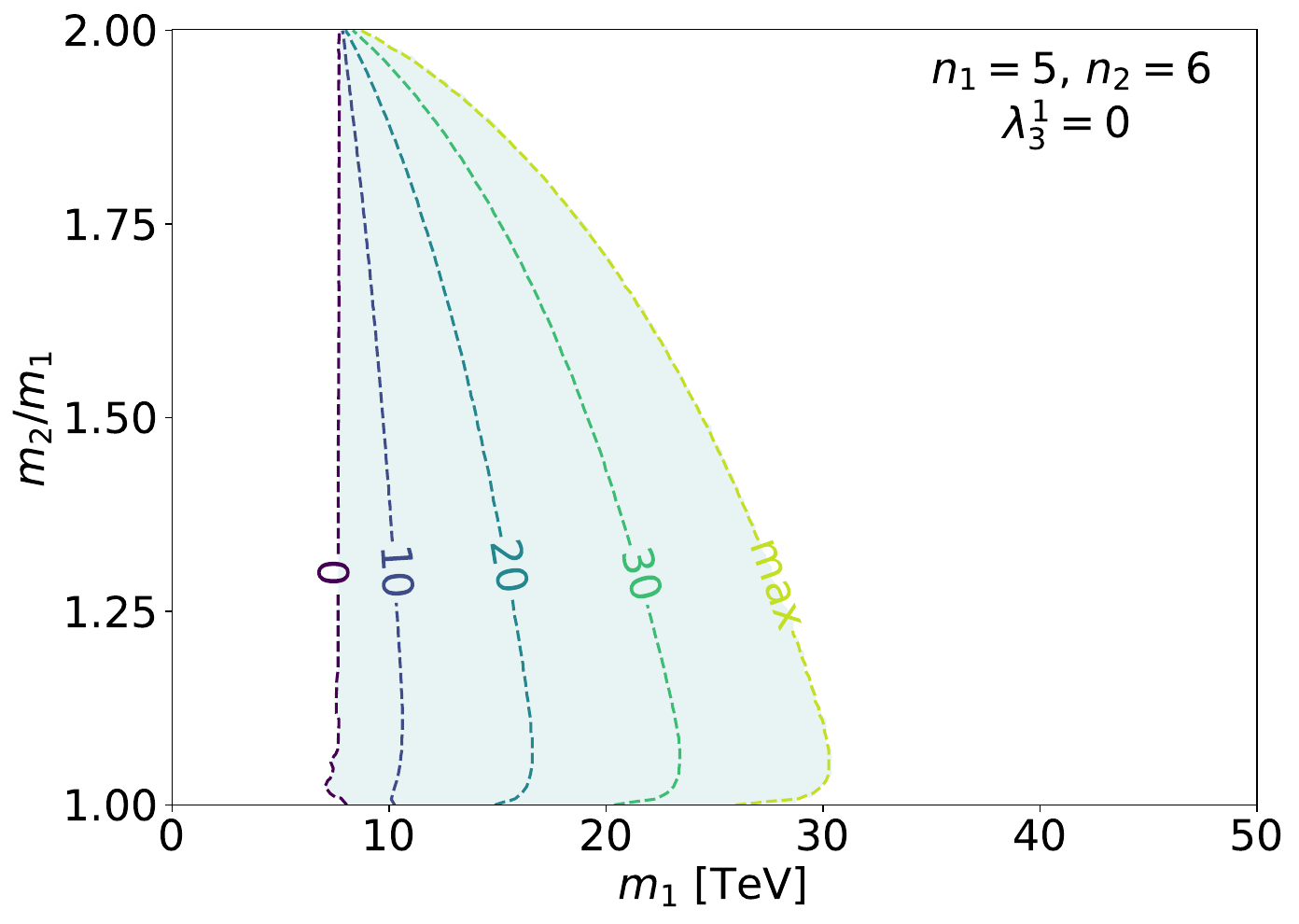}
        \label{fig:C56a}
    \end{subfigure}    
    \begin{subfigure}{0.49\textwidth}
        \centering
        \caption{}
        \includegraphics[width=\textwidth]{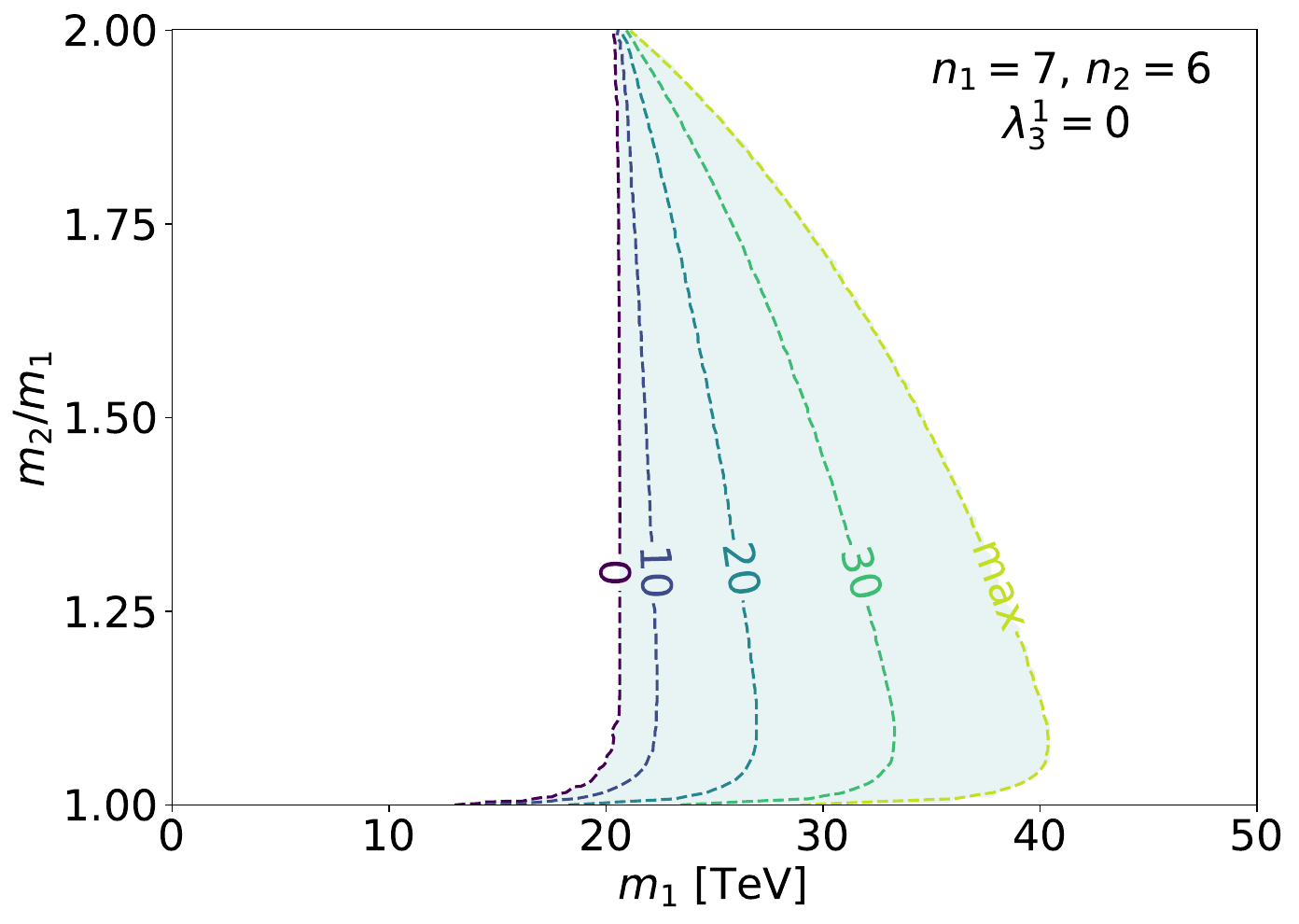}
        \label{fig:C76a}
    \end{subfigure}    
    \caption{Parameter space as a function of $m_1$ and $m_2/m_1$ for different $n_1$ and $n_2$ when the parameter $\lambda_3^1$ is set to 0. All other parameters are set to 0. Contours of $\lambda_1$ are shown. The max value corresponds to the unitarity bound of Eq.~\eqref{eq:UnitarityFinal}. The green region is allowed.}
    \label{fig:Ca}
\end{figure}

\begin{figure}[t!]
    \centering
    \begin{subfigure}{0.49\textwidth}
        \centering
        \caption{}
        \includegraphics[width=\textwidth]{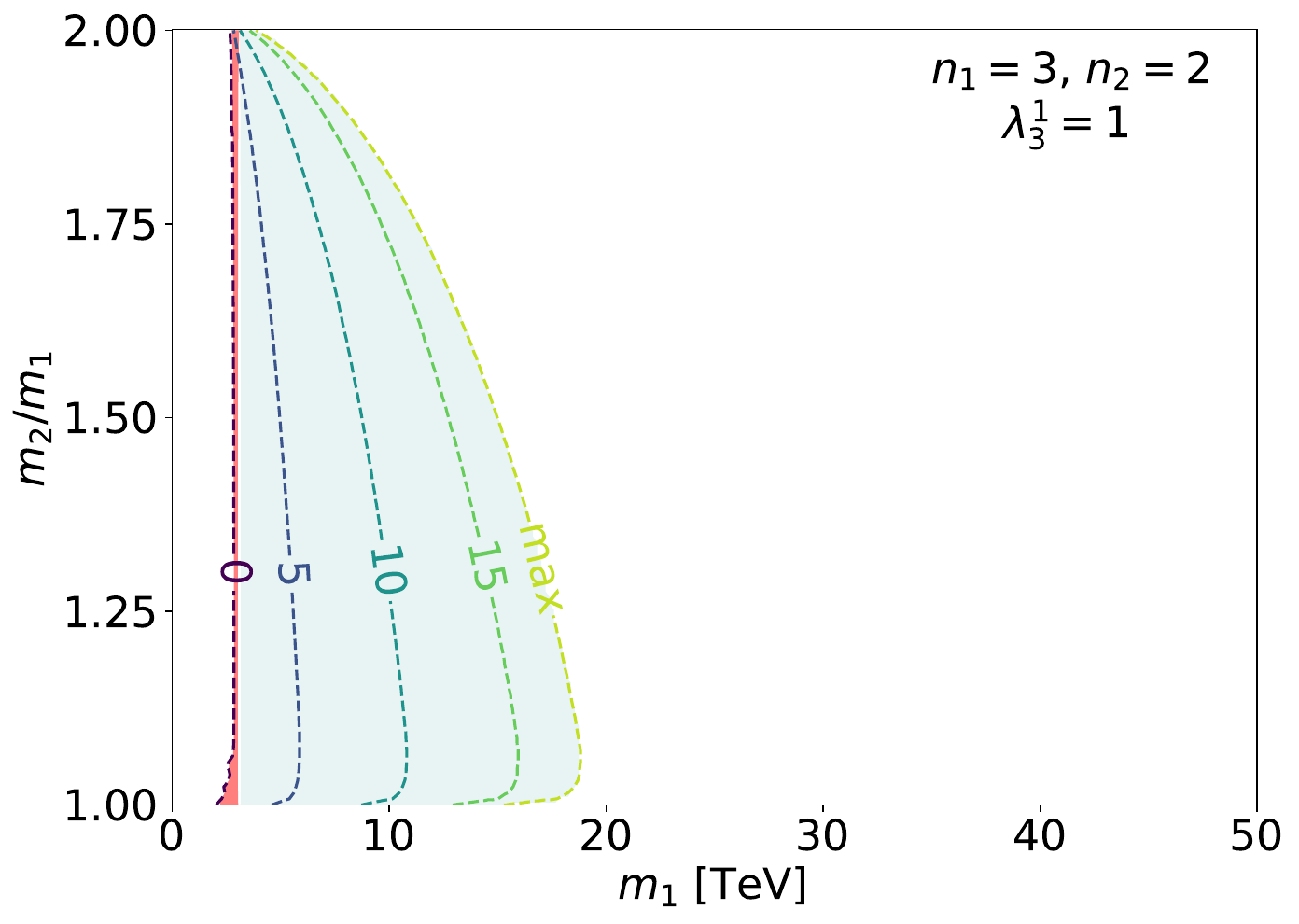}
        \label{fig:C32b}
    \end{subfigure}    
    \begin{subfigure}{0.49\textwidth}
        \centering
        \caption{}
        \includegraphics[width=\textwidth]{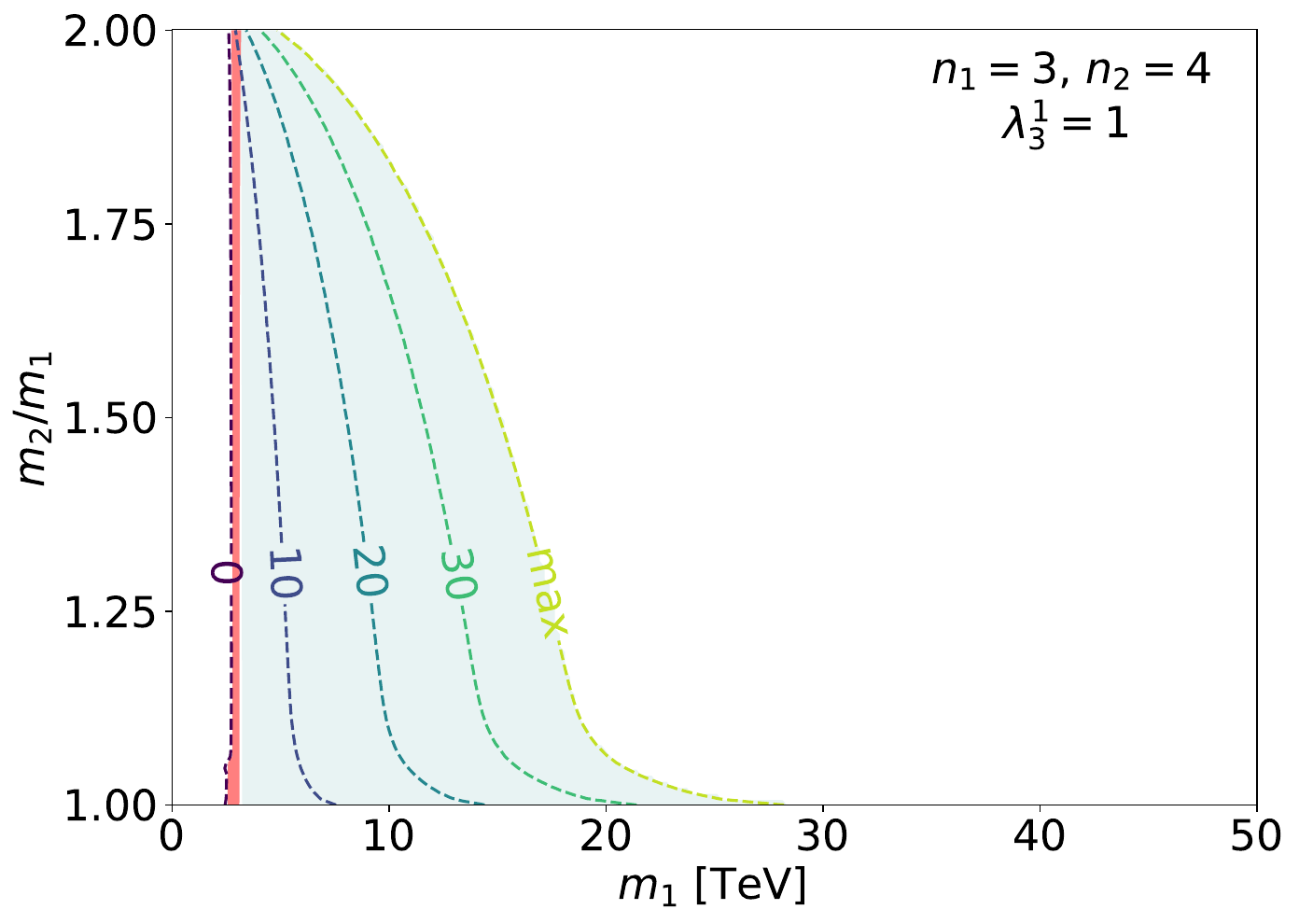}
        \label{fig:C34b}
    \end{subfigure}
    \begin{subfigure}{0.49\textwidth}
        \centering
        \caption{}
        \includegraphics[width=\textwidth]{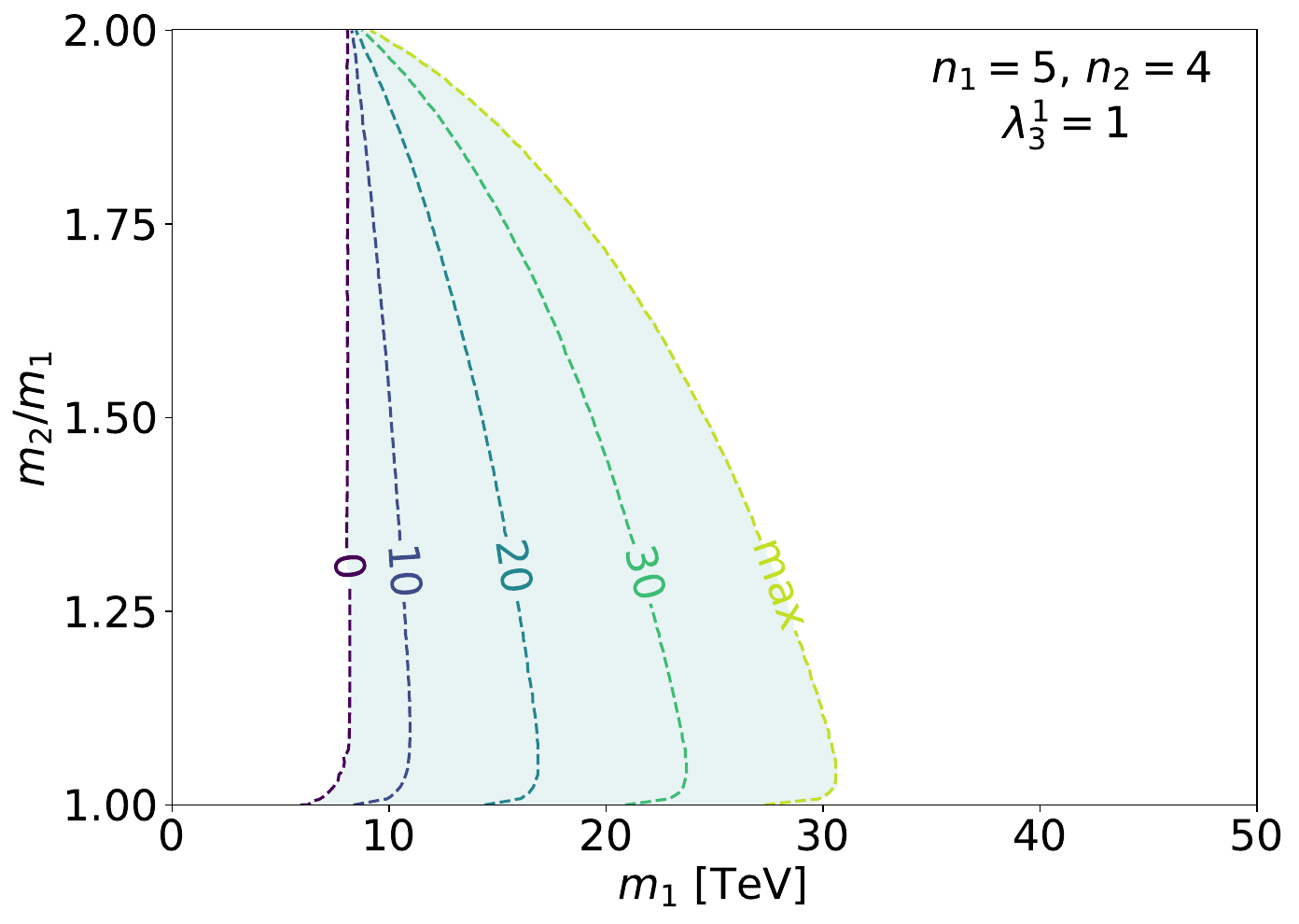}
        \label{fig:C54b}
    \end{subfigure}    
    \begin{subfigure}{0.49\textwidth}
        \centering
        \caption{}
        \includegraphics[width=\textwidth]{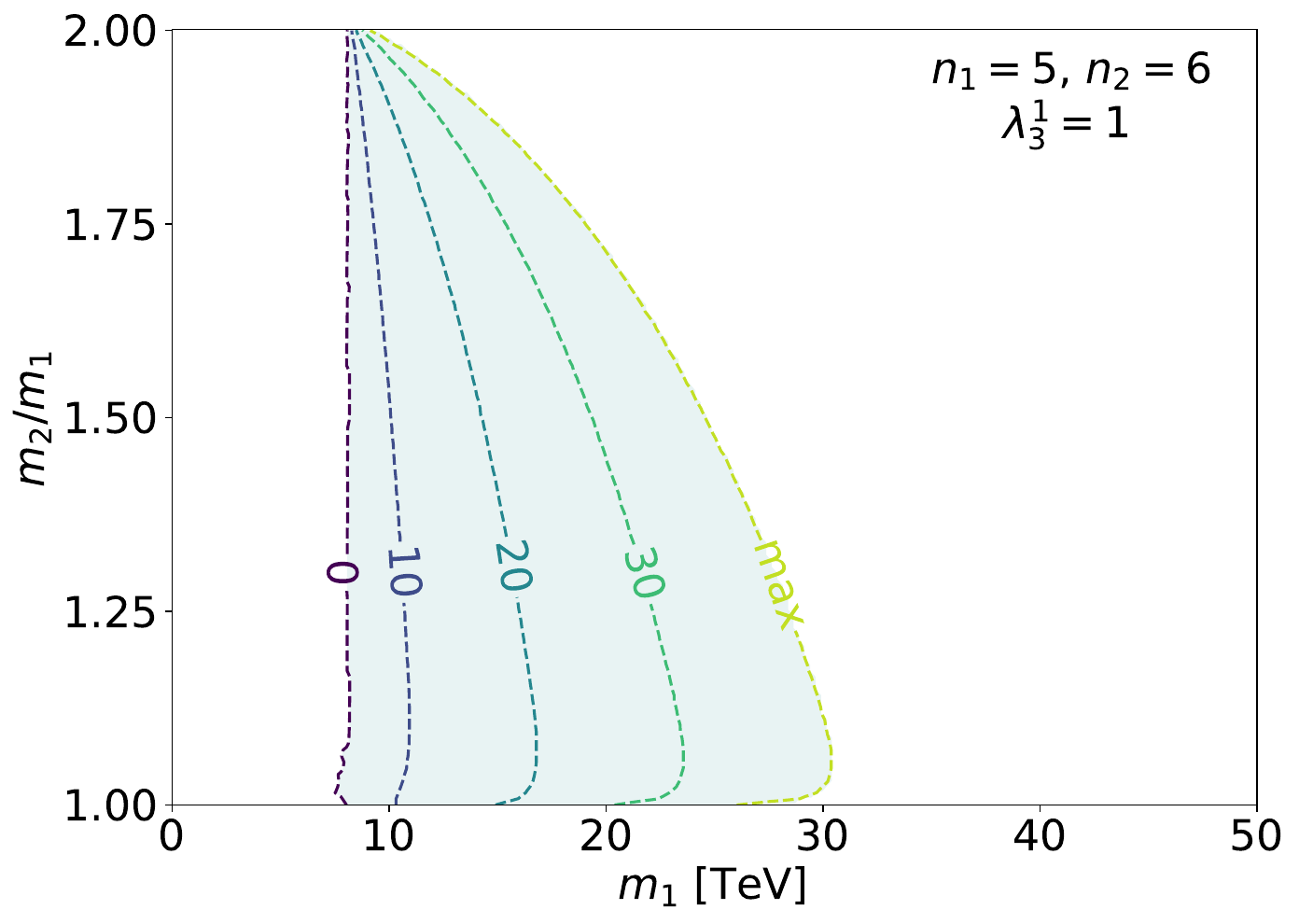}
        \label{fig:C56b}
    \end{subfigure}    
    \begin{subfigure}{0.49\textwidth}
        \centering
        \caption{}
        \includegraphics[width=\textwidth]{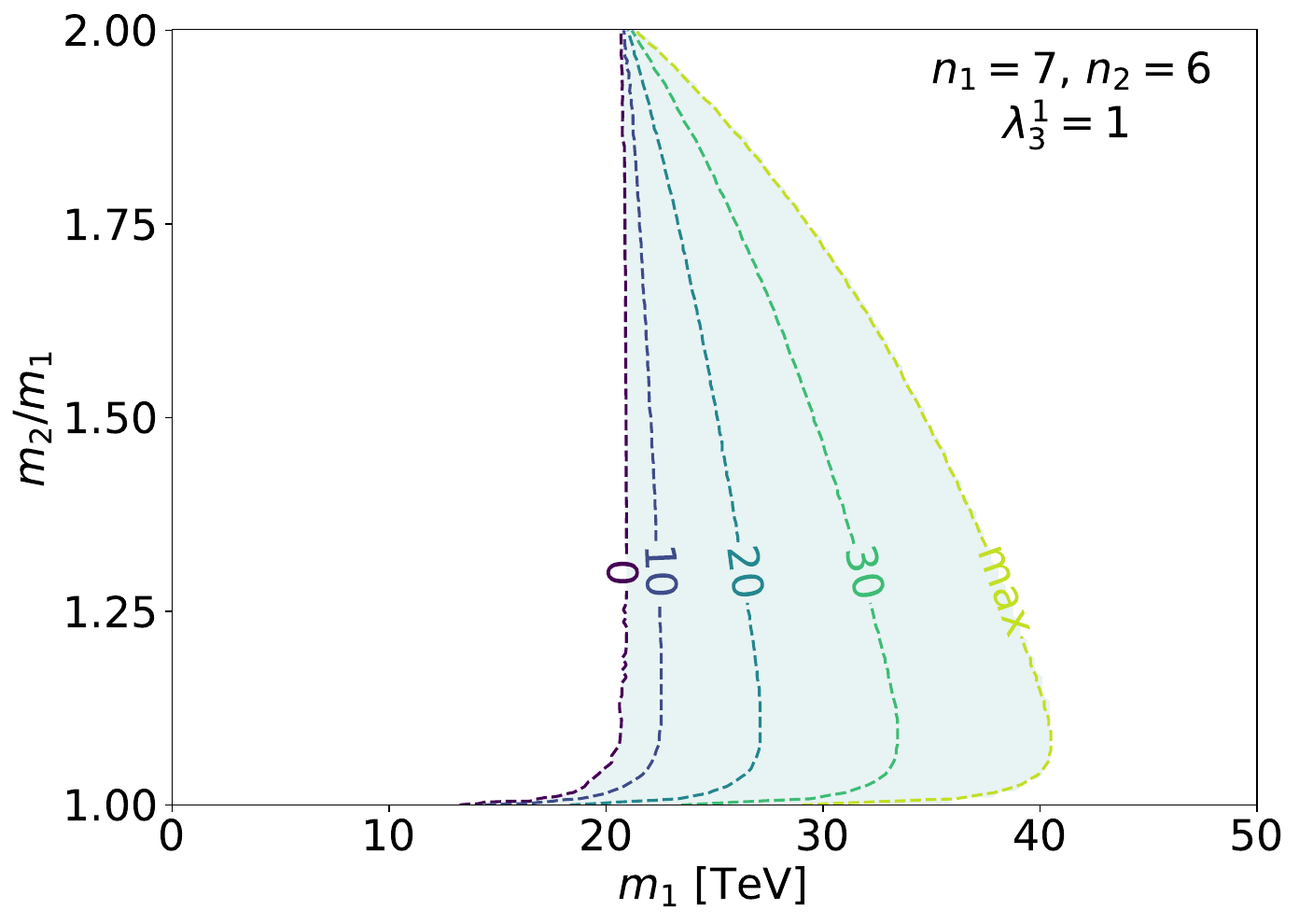}
        \label{fig:C76b}
    \end{subfigure}    
    \caption{Same as Fig.~\ref{fig:Ca} but for $\lambda_3^1 = 1$. The green region is allowed and the red region is forbidden by direct detection experiments.}
    \label{fig:Cb}
\end{figure}

\begin{figure}[t!]
    \centering
    \begin{subfigure}{0.49\textwidth}
        \centering
        \caption{}
        \includegraphics[width=\textwidth]{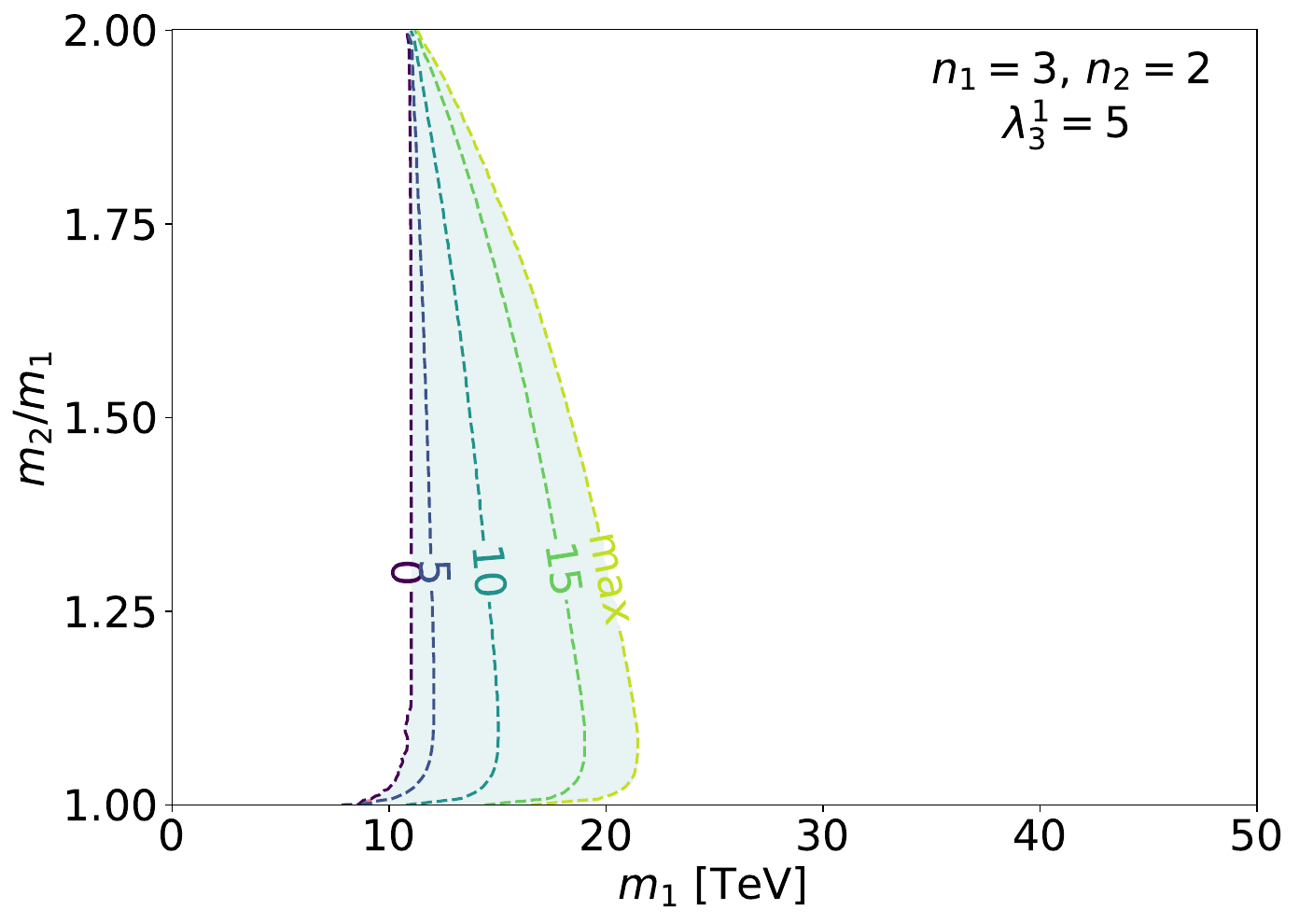}
        \label{fig:C32c}
    \end{subfigure}    
    \begin{subfigure}{0.49\textwidth}
        \centering
        \caption{}
        \includegraphics[width=\textwidth]{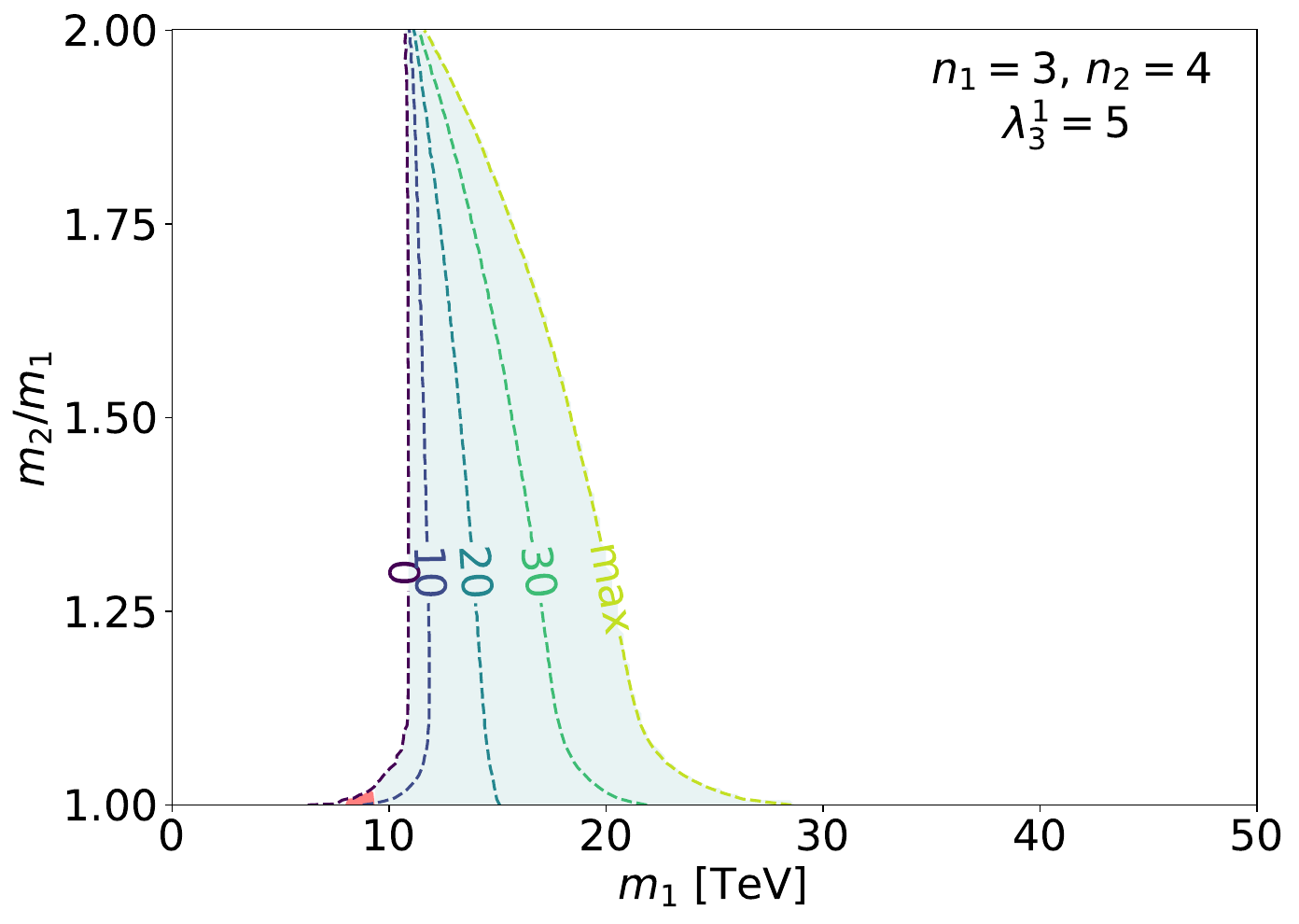}
        \label{fig:C34c}
    \end{subfigure}
    \begin{subfigure}{0.49\textwidth}
        \centering
        \caption{}
        \includegraphics[width=\textwidth]{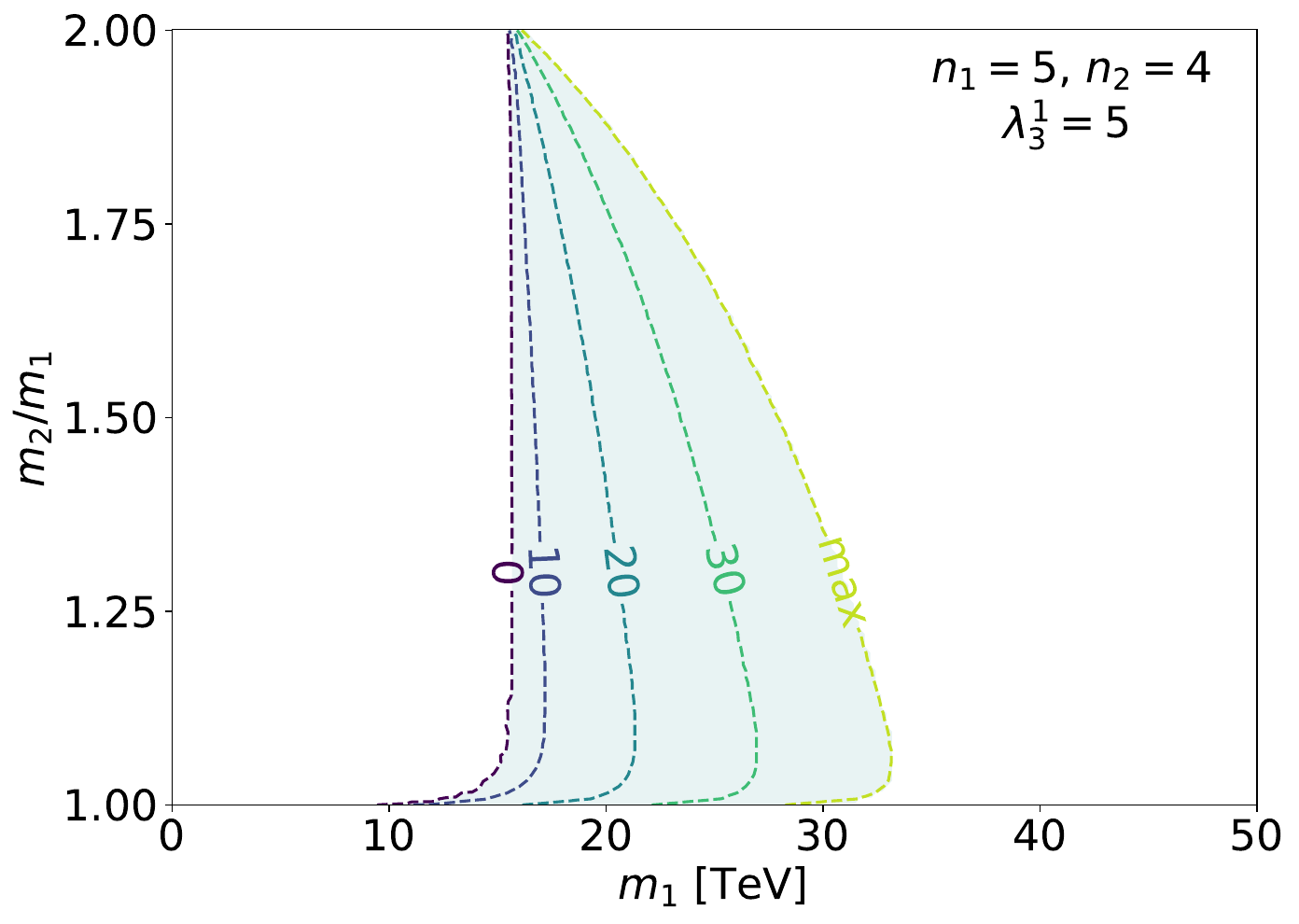}
        \label{fig:C54c}
    \end{subfigure}    
    \begin{subfigure}{0.49\textwidth}
        \centering
        \caption{}
        \includegraphics[width=\textwidth]{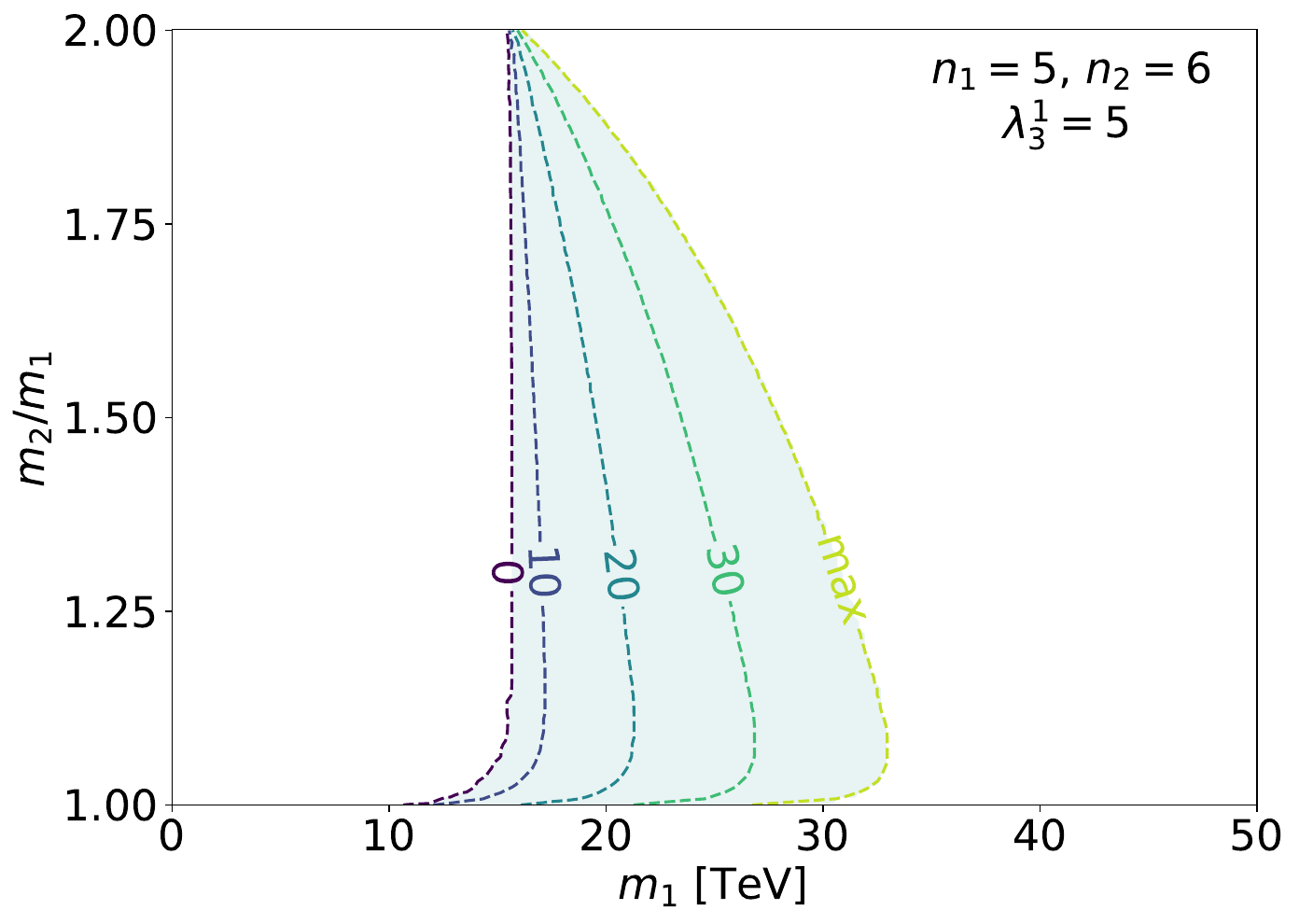}
        \label{fig:C56c}
    \end{subfigure}    
    \begin{subfigure}{0.49\textwidth}
        \centering
        \caption{}
        \includegraphics[width=\textwidth]{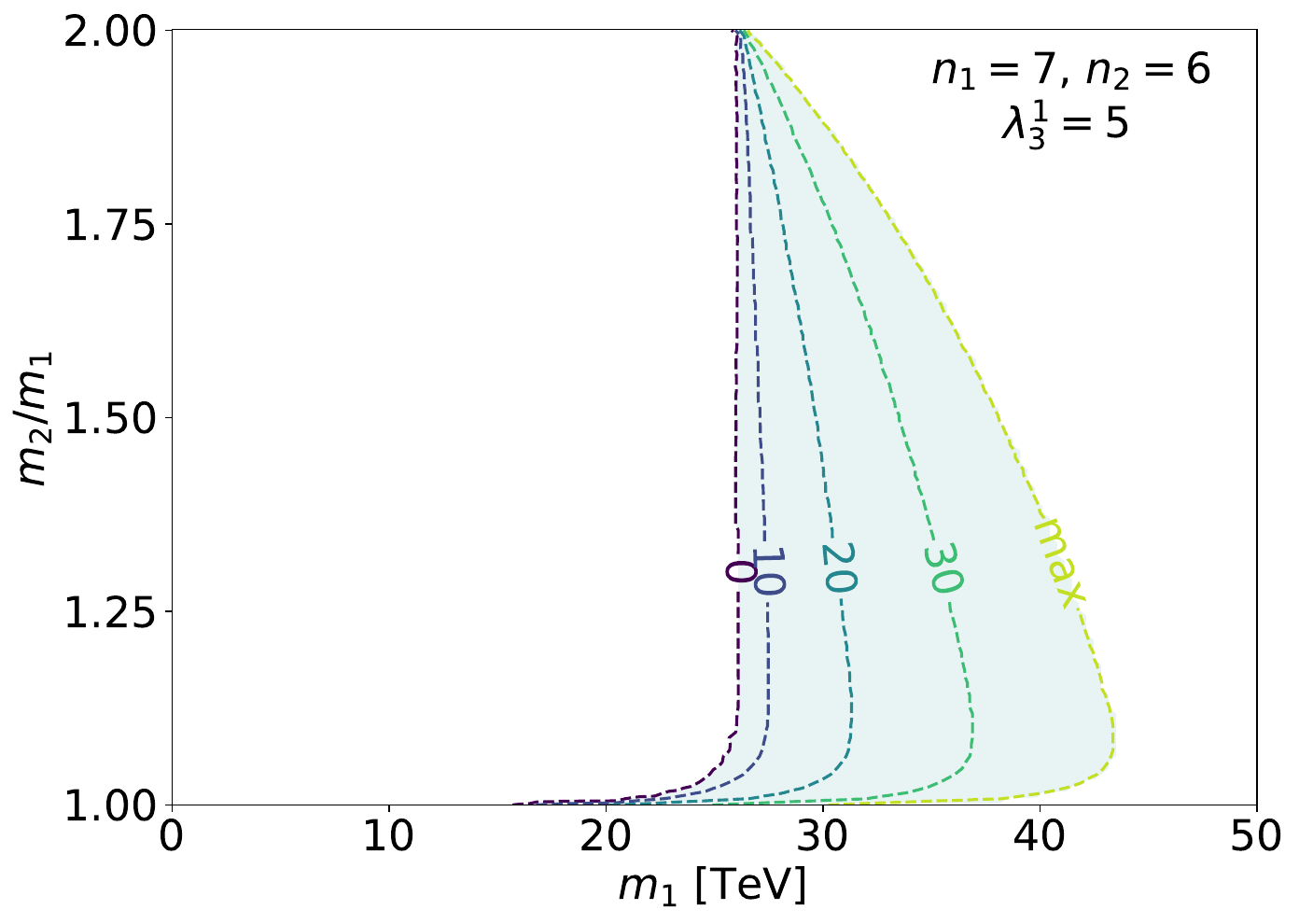}
        \label{fig:C76c}
    \end{subfigure}    
    \caption{Same as Fig.~\ref{fig:Ca} but for $\lambda_3^1 = 5$. The green region is allowed and the red region is forbidden by direct detection experiments.}
    \label{fig:Cc}
\end{figure}

Figs. \ref{fig:Ca}, \ref{fig:Cb} and \ref{fig:Cc} show the allowed region in the space of $m_1$ and $m_2/m_1$ for different combinations of $n_1$ and $n_2$ when $\lambda_3^1$ is set respectively to 0, 1 and 5. Combinations of larger multiplets would be in conflict with the unitarity bounds of Ref.~\cite{Hally:2012pu}.\footnote{An additional requirement on the absence of a Landau pole below a specific energy could also impose an upper limit on $n_1$ and $n_2$, see, e.g., Ref.~\cite{Hamada:2015bra}.} We also ignore the case of $n_1 = 1$ and $n_2 = 2$, as the presence of a singlet is not really aligned with the goals of this paper, is somewhat qualitatively different, and has previously been studied extensively (see, e.g., Ref.~\cite{Belanger:2014bga}). In each case, the parameter $\lambda_1$ is adjusted to reproduce the correct relic abundance. Contours of this parameter are also shown. All other parameters are set to zero, as their impact on the dark matter abundance would be small or as they are restricted to be small by other constraints. It is however assumed that $\lambda_5$ has a small non-zero value for $\phi_1$ and $\phi_2$ to maintain chemical equilibrium, but not sufficiently large to noticeably alter the relic abundance. The ratio $m_2/m_1$ is always taken as larger than 1, as this would otherwise be in conflict with direct detection experiments. The lower limit on $m_1$ comes from the impossibility of adjusting $\lambda_1$ to reproduce the relic abundance or possibly direct detection constraints when $\lambda_3^1$ is non-zero. The upper limit comes from the unitarity constraint of Eq.~\eqref{eq:UnitarityFinal}.

As can be seen, the presence of semi-annihilation processes allows $m_1$ to span a range of sometimes more than an order of magnitude. In scenarios with a $\mathbb{Z}_2$ symmetries, direct detection constraints often force the multiplet masses to be in a very narrow range, as the only freedom to change the abundance comes from heavily constrained interactions with the Higgs boson. The allowed range does tend to relatively shrink as the size of the multiplets increases. As was to be expected, increasing $m_2/m_1$ typically reduces the impact of semi-annihilation processes. The allowed range is especially narrow for values of $m_2/m_1$ larger than 2, as the process $\phi_1 \phi_1 \to \phi_2^\dagger H^\dagger$ becomes kinematically forbidden. However, the plots show that semi-annihilation processes can still be very efficient even when $m_1$ and $m_2$ differ considerably. As such, semi-annihilation does not require careful fine-tuning between the masses to be efficient, as is sometimes the case with coannihilation~\cite{Griest:1990kh}. Increasing the value of $\lambda_3^1$ results in some of the lower mass region being excluded by direct detection constraints for smaller multiplets, but shifts the upper limit on $m_1$ toward larger values. The behaviour around $m_2/m_1 = 1$ is non-trivial and comes from a combination of efficient coannhihilation and Sommerfeld enhancement with different signs of $R$, $R'_1$ and $R'_2$. In all plots, the dark matter candidate is mostly $\phi_1$-like and with a mass very close to $m_1$.

\section{Conclusion}\label{Sec:Conclusion}
The goal of this work is to determine which models of inert multiplets can lead to efficient semi-annihilation while at the same time being compatible with known constraints and naturally avoiding stable charged particles.

We found that semi-annihilation cannot be efficient in the case of a single multiplet. For two multiplets, there exists only one category of models that can lead to efficient semi-annihilation while avoiding strong constraints from direct detection experiments and stable charged particles at tree level. Other models exist, but must rely on loop corrections to avoid stable charged particles, meet certain conditions to avoid direct detection constraints, or both. For smaller multiplets, semi-annihilation allows the masses of the scalars to reproduce the correct relic abundance over a range of sometimes more than an order of magnitude. This relative range shrinks as the sizes of the multiplets increase.

As should be obvious, this paper only constitutes a first study of semi-annihilation of inert multiplets and many questions remain unanswered. We focused on what we considered to be the most attractive model, but some of the other options presented are far from forbidden. Furthermore, the model of this paper can indeed lead to the lightest particle being neutral at tree level, but meeting this condition is somewhat difficult numerically and it is not that hard for other models to avoid stable charged particles when loop corrections are taken into account. The other models would certainly warrant further studies. In addition, collider constraints were not considered in this work, as they are currently not very stringent and can easily be avoided. Potential signatures include charged particles stable on collider scales, disappearing tracks and missing energy. Prospects at future colliders already exist for inert multiplets \cite{Kadota:2018lrt, Zeng:2019tlw, Chiang:2020rcv, Liu:2020dok, Bottaro:2021snn, Bottaro:2022one}, but it would be interesting to determine how they are affected by the presence of semi-annihilation processes.

More importantly, we did not perform a dedicated analysis of indirect detection constraints. As is well known, indirect detection experiments now impose considerable constraints on inert multiplets, often to the point that standard $\mathbb{Z}_2$ scenarios struggle to accommodate the measured abundance~\cite{Arakawa:2021vih, Katayose:2021mew}. Performing a dedicated study of indirect detection constraints would entail two rather difficult tasks. First, one should compute a limit on the indirect detection cross section for semi-annihilation. Even though some work already exists on the subject~\cite{Cai:2016hne, Queiroz:2019acr}, we are unaware of any existing work general enough to cover all signals from inert multiplet semi-annihilation. Second, one would need to carefully compute the Sommerfeld-enhanced cross section for different astrophysical sources.

However, it is not necessary to perform such tasks to already tell that semi-annihilation can alleviate the tension with indirect detection experiments. The presence of semi-annihilation processes allows the dark matter to be considerably heavier. This by itself suppresses the constraints from annihilation to Standard Model particles. Semi-annihilation processes themselves can lead to an indirect detection signal, but the kinematics should lead to reduced limits. Most importantly, there exist combinations of gauge numbers for which semi-annihilation can actually be suppressed by the Sommerfeld enhancement. This is namely the case for $n_1 = 3$ and $n_2 = 4$. Such cases could therefore have negligible signals from indirect detection experiments. Many possibilities also exist when one is willing to go beyond the simple model studied in this paper. As such, it could very well be that semi-annihilation of inert multiplets will become an increasingly attractive possibility as constraints from indirect detection experiments keep improving.

\acknowledgments
C.-W.~C would like to thank the High Energy Physics Theory Group at the University of Tokyo for their hospitality during his visit when part of this work was finished. This work was supported by the National Science and Technology Council under Grant No. NSTC-111-2112-M-002-018-MY3, the Ministry of Education (Higher Education Sprout Project NTU-112L104022), and the National Center for Theoretical Sciences of Taiwan.

\appendix

\section{$SU(2)$ tensors}\label{Sec:SU2}
We present in this section the different $SU(2)$ tensors used in the paper. Although not a very enlightening discussion, not explicitly stating them would lead to ambiguities on the meaning of certain coefficients as different bases exist in the literature.

\subsection{One multiplet case}\label{sSec:SU2OneMultiplet}
For the model with a single multiplet, the $A_{abc}$ tensor is written as
\begin{equation}\label{eq:Model1A}
  A_{abc} = \sum_{M=-J}^J C^{JM}_{j_1 m_1 j_2 m_2} C^{00}_{J M j_3 m_3},
\end{equation}
where $C^{JM}_{j_1 m_1 j_2 m_2} = \langle j_1 j_2 m_1 m_2 |J M \rangle$ are the Clebsch-Gordan coefficients and
\begin{equation}\label{eq:Model1Ab}
  \begin{aligned}
    j_1 &= \frac{n - 1}{2},      & j_2 &= \frac{n - 1}{2},      & j_3 &= \frac{n - 1}{2},      & J &= \frac{n - 1}{2},\\ 
    m_1 &= \frac{n + 1 - 2a}{2}, & m_2 &= \frac{n + 1 - 2b}{2}, & m_3 &= \frac{n + 1 - 2c}{2}.
  \end{aligned}
\end{equation}
For the sake of convenience, the $B^r_{abcd}$ tensors are written as
\begin{equation}\label{eq:Model1B}
  B^1_{abcd} = \delta_{ab} \delta_{cd}, \quad B^2_{abcd} = \tau^e_{ab} T^e_{cd},
\end{equation}
where $\tau^M_{ab}$ and $T^M_{cd}$ are respectively the 2- and $n$-dimensional $SU(2)$ generators. The $C^r_{abcd}$ tensors are given by
\begin{equation}\label{eq:Model1C}
  C^r_{abcd} = \sum_{M=-J}^J C^{JM}_{j_1 m_1 j_2 m_2} C^{JM}_{j_3 m_3 j_4 m_4},
\end{equation}
where
\begin{equation}\label{eq:Model1Cb}
  \begin{aligned}
    J &= n + 1 - 2r,\\
    j_1 &= \frac{n - 1}{2},      & j_2 &= \frac{n - 1}{2},      & j_3 &= \frac{n -1}{2},       & j_4 &= \frac{n -1}{2},\\ 
    m_1 &= \frac{n + 1 - 2a}{2}, & m_2 &= \frac{n + 1 - 2b}{2}, & m_3 &= \frac{n + 1 - 2c}{2}, & m_4 &= \frac{n + 1 - 2d}{2},
  \end{aligned}
\end{equation}
where $r \in \{1, 2, \ldots, \lfloor (n + 1)/2 \rfloor\}$.

\subsection{Two multiplets case}\label{sSec:SU2TwoMultiplets}
For the model with two multiplets, the $A_{abcd}$ tensors are given by
\begin{equation}\label{eq:Model2A}
  A_{abcd} = \sum_{M, M'=-J}^J C^{JM}_{j_1 m_1 j_2 m_2} C^{JM'}_{j_3 m_3 j_4 m_4} C^{00}_{J M J M'},
\end{equation}
where
\begin{equation}\label{eq:Model2Ab}
  \begin{aligned}
    R   &= 4\left\lfloor\frac{n_2}{4}\right\rfloor  + 1, & J &= \frac{R - 1}{2},\\
    j_1 &= \frac{n_1 - 1}{2},      & j_2 &= \frac{n_1 - 1}{2},      & j_3 &= \frac{n_2 -1}{2},       & j_4 &= \frac{1}{2},\\ 
    m_1 &= \frac{n_1 + 1 - 2a}{2}, & m_2 &= \frac{n_1 + 1 - 2b}{2}, & m_3 &= \frac{n_2 + 1 - 2c}{2}, & m_4 &= \frac{3 - 2d}{2}.
  \end{aligned}
\end{equation}
The $B_{abc}$ tensor is given by
\begin{equation}\label{eq:Model2B}
  B_{abc} = \sum_{M=-J}^J C^{JM}_{j_1 m_1 j_2 m_2} C^{00}_{J M j_3 m_3},
\end{equation}
where
\begin{equation}\label{eq:Model2Bb}
  \begin{aligned}
    j_1 &= \frac{n_1 - 1}{2},      & j_2 &= \frac{n_1 - 1}{2},      & j_3 &= \frac{n_1 - 1}{2},      & J &= \frac{n_1 - 1}{2},\\ 
    m_1 &= \frac{n_1 + 1 - 2a}{2}, & m_2 &= \frac{n_1 + 1 - 2b}{2}, & m_3 &= \frac{n_1 + 1 - 2c}{2}.
  \end{aligned}
\end{equation}
The $C^r_{abcd}$ and $D^r_{abcd}$ tensors are written as
\begin{equation}\label{eq:Model2CD}
  \begin{aligned}
    C^1_{abcd} &= \delta_{ab} \delta_{cd}, \quad C^2_{abcd} = \tau^e_{ab} T^e_{cd},\\
    D^1_{abcd} &= \delta_{ab} \delta_{cd}, \quad D^2_{abcd} = \tau^e_{ab} \hat{T}^e_{cd},
  \end{aligned}
\end{equation}
where $T^M_{cd}$ and $\hat{T}^M_{cd}$ are respectively the $SU(2)$ generators of dimension $n_1$ and $n_2$. The $E_{abc}$ tensor is given by
\begin{equation}\label{eq:Model2E}
  E_{abc} = C^{JM}_{j_1 m_1 j_2 m_2},
\end{equation}
where
\begin{equation}\label{eq:Model2Eb}
  \begin{aligned}
    j_1 &= \frac{n_2 - 1}{2},      & j_2 &= \frac{1}{2},      & J &= \frac{n_1 - 1}{2},\\ 
    m_1 &= \frac{n_2 + 1 - 2b}{2}, & m_2 &= \frac{3 - 2c}{2}, & M &= \frac{n_1 + 1 - 2a}{2}.
  \end{aligned}
\end{equation}
The $F^r_{abcd}$ tensor is written as
\begin{equation}\label{eq:Model2F}
  F^r_{abcd} = \sum_{M=-J}^J C^{JM}_{j_1 m_1 j_2 m_2} C^{JM}_{j_3 m_3 j_4 m_4},
\end{equation}
where
\begin{equation}\label{eq:Model2Fb}
  \begin{aligned}
    J &= n_1 + 1 - 2r,\\
    j_1 &= \frac{n_1 - 1}{2},      & j_2 &= \frac{n_1 - 1}{2},      & j_3 &= \frac{n_1 -1}{2},       & j_4 &= \frac{n_1 -1}{2},\\ 
    m_1 &= \frac{n_1 + 1 - 2a}{2}, & m_2 &= \frac{n_1 + 1 - 2b}{2}, & m_3 &= \frac{n_1 + 1 - 2c}{2}, & m_4 &= \frac{n_1 + 1 - 2d}{2},
  \end{aligned}
\end{equation}
where $r \in \{1, 2, \ldots, \lfloor (n_1 + 1)/2 \rfloor\}$. The $G^r_{abcd}$ tensor is written as
\begin{equation}\label{eq:Model2G}
  G^r_{abcd} = \sum_{M=-J}^J C^{JM}_{j_1 m_1 j_2 m_2} C^{JM}_{j_3 m_3 j_4 m_4},
\end{equation}
where
\begin{equation}\label{eq:Model2Gb}
  \begin{aligned}
    J &= n_2 + 1 - 2r,\\
    j_1 &= \frac{n_2 - 1}{2},      & j_2 &= \frac{n_2 - 1}{2},      & j_3 &= \frac{n_2 -1}{2},       & j_4 &= \frac{n_2 -1}{2},\\ 
    m_1 &= \frac{n_2 + 1 - 2a}{2}, & m_2 &= \frac{n_2 + 1 - 2b}{2}, & m_3 &= \frac{n_2 + 1 - 2c}{2}, & m_4 &= \frac{n_2 + 1 - 2d}{2},
  \end{aligned}
\end{equation}
where $r \in \{1, 2, \ldots, \lfloor (n_2 + 1)/2 \rfloor\}$. Finally, the $H^r_{abcd}$ tensor is written as
\begin{equation}\label{eq:Model2H}
  H^r_{abcd} = \sum_{M=-J}^J C^{JM}_{j_1 m_1 j_2 m_2} C^{JM}_{j_3 m_3 j_4 m_4},
\end{equation}
where
\begin{equation}\label{eq:Model2Hb}
  \begin{aligned}
    J &= \frac{n_1 + n_2 - 2r}{2},\\
    j_1 &= \frac{n_1 - 1}{2},      & j_2 &= \frac{n_2 - 1}{2},      & j_3 &= \frac{n_1 -1}{2},       & j_4 &= \frac{n_2 -1}{2},\\ 
    m_1 &= \frac{n_1 + 1 - 2a}{2}, & m_2 &= \frac{n_2 + 1 - 2b}{2}, & m_3 &= \frac{n_1 + 1 - 2c}{2}, & m_4 &= \frac{n_2 + 1 - 2d}{2},
  \end{aligned}
\end{equation}
where $r \in \{1, 2, \ldots, \text{min}(n_1, n_2) \}$.

\bibliography{biblio}
\bibliographystyle{utphys}

\end{document}